\documentclass[11pt,a4paper]{article}

\usepackage[utf8x]{inputenc}
\usepackage[T1]{fontenc}
\usepackage{color}
\usepackage{amssymb}
\usepackage{amsmath,amsfonts}
\usepackage{graphicx}
\usepackage{mathtools}
\usepackage{ragged2e}
\usepackage{psfrag}
\usepackage{relsize}
\usepackage{upgreek}
\usepackage{booktabs}

\usepackage{bm}
\usepackage{mathrsfs}
\usepackage{graphicx}
\usepackage{verbatim} 
\usepackage[english]{babel}
\usepackage[hypertexnames=false]{hyperref} 
\usepackage{url}
\hypersetup{
citecolor=red,
colorlinks=true,
filecolor=red,
linkcolor=blue,
linktocpage=true,
urlcolor=blue
} 
\usepackage[titletoc,toc,title]{appendix}
\numberwithin{equation}{section}
\usepackage{cleveref}
\usepackage{cite}
\usepackage{epsfig}
\usepackage{float}
\usepackage{enumitem}
\usepackage[font={footnotesize,it}]{caption}
\usepackage{cases}
\usepackage{tikz,empheq}

\usepackage{tocloft}
\newcommand{\spll}{{/\kern-0.2em/}}
\newcommand\trick[1]{}
\newcommand{\be}{\begin{equation}} 
\newcommand{\ee}{\end{equation}}
\newcommand{\eq}[1]{(\ref{#1})}
\newcommand{\bit}{\begin{itemize}}  \newcommand{\eit}{\end{itemize}}
\newcommand{\ben}{\begin{enumerate}}  \newcommand{\een}{\end{enumerate}}

\newcommand{\Tr}{\operatorname{Tr}}
\newcommand{\dd}{\mathrm d}
\newcommand{\ii}{\mathrm i}

\def\bea{\begin{eqnarray}}
\def\eea{\end{eqnarray}}

\def\Tr{{\rm Tr}}
\def\dd{\mathrm d}
\def\k{\kappa}
 
 \def\n{\nu}

\def\s{\sigma}

 \def\cN{{\cal N}}

\newcommand{\nn}{{\nonumber}}

\def\uno{\mbox{1 \kern-.59em {\rm l}}}

\def\dag{\dagger}

\def\bcomment#1{}

 \def\ii{{\rm i}}

\def\IR{\relax{\rm I\kern-.18em R}}

\long\def\symbolfootnote[#1]#2{\begingroup%
\def\thefootnote{\fnsymbol{footnote}}\footnote[#1]{#2}\endgroup}
\newcommand{\nthu}{{\it Department of Physics, National Tsing-Hua University,
    Hsinchu 30013, Taiwan}}

\newcommand{\ctc}{{\it
Center for Theory-Computation-Data Science Research, 
National Tsing-Hua University, Hsinchu 30013, Taiwan}}

\newcommand{\ncts}{{\it Physics Division,
    National Center for Theoretical Sciences, Taipei 10617, Taiwan}}

\begin{document}
\begin{center}
\vspace{20pt}

\thispagestyle{empty}
              {\Large \bf Cosmological Constant Screening
                from  Quantum Horizon Chaos
                \\[1.0ex]
{\normalfont\normalsize\itshape
Dedicated to the memory of Yoshihisa Kitazawa}
              }

\vspace{25pt}

Chong-Sun Chu

\vspace{0.2cm}              

\vspace{5pt}\nthu\\
\vspace{5pt}\ctc\\
\vspace{5pt}\ncts

\vspace{1cm}

\begin{abstract}

We propose a microscopic mechanism for screening the cosmological
constant through quantum chaos of the horizon degrees of freedom. We
show that in a recently proposed quantum-horizon matrix model
\cite{Chu:2024qil}, the Schwarzschild--de Sitter junction completion
\cite{Chu:2026kqt} supports a gravitationally dressed soft mode that
approaches the massless minimally coupled scalar of de Sitter
gravity. Interactions among the microscopic horizon degrees of freedom
provide a mechanism for force relaxation. Assuming a finite-memory
force spectrum with nonzero regular low-frequency weight, the
gravitationally dressed soft mode undergoes Brownian diffusion after
momentum relaxation. The resulting bosonic coarse-grained entropy
supplements the background entropy arising from fermionic parton
microstates. We propose that their sum defines the total thermodynamic
entropy, which can be identified with the Gibbons--Hawking entropy of
the backreacted causal horizon.  Entropy generation during the
diffusion regime then leads to infrared screening of the effective
cosmological constant, with a universal entropy--variance coefficient
agreeing with that of Kitazawa and collaborators. Radiative lifting of
the soft mode provides a possible mechanism for terminating the
screening regime.

\end{abstract}

\end{center}

\newpage
\setcounter{footnote}{0}
%\setcounter{page}{1}
%%%%%%%%%%%%%%%%%%%%%%%%%%%%%%%

\tableofcontents
%%%%%%%%%%%%%%%%%%%%%%%%%%%%%%%
\section{Introduction}

Quantum fluctuations in de Sitter space can accumulate over long times
and produce significant infrared effects. In the stochastic approach
pioneered by Starobinsky \cite{Starobinsky:1982ee,Starobinsky:1986fx},
short-wavelength quantum fluctuations are continuously transferred
into the long-wavelength sector as their physical wavelengths cross
the horizon scale. The coarse-grained long-wavelength field
consequently receives quantum kicks and acquires an effective Langevin
dynamics, together with an associated Fokker--Planck description.
The quantum foundation of this stochastic description has been
developed systematically through reduced-density-matrix and
open-quantum-system approaches, which derive the effective dynamics by
integrating out short-wavelength degrees of freedom
\cite{Burgess:2014eoa,Collins:2017haz,Li:2025azq}. The resulting
stochastic theory captures the secular growth of long-wavelength
scalar fluctuations.

Recent developments have further clarified this open-quantum-system
picture. Unlike a fixed system--environment separation, stochastic
inflation involves a time-dependent coarse graining as modes
continually cross the horizon and enter the infrared sector
\cite{Christie:2026dwx}. Open-effective-field-theory analyses have
further distinguished this stochastic evolution from conventional
Wilsonian renormalization and examined the emergence of Markovian
dynamics and possible non-Markovian corrections \cite{Li:2026lwl}.

It is natural to expect an analogous mechanism for gravity itself.
The slowly varying de Sitter geometry should interact with faster
quantum-gravitational degrees of freedom, whose coarse-grained
fluctuations act as a stochastic force on the long-wavelength
geometry. Polyakov argued that infrared quantum effects may
destabilize de Sitter space and lead to screening of the cosmological
constant \cite{Polyakov:2007mm,Polyakov:2009nq}.
Stochastic approaches to de Sitter gravity provide a related
perspective. In particular, Kitazawa and collaborators have argued
that the conformal zero mode exhibits secular Brownian growth and an
associated screening of the effective cosmological constant
\cite{Kitamoto:2019rij,Kitazawa:2020oaf,
  Kitazawa:2024deb}.  The physical interpretation of secular infrared
effects in de Sitter gravity has, however, been debated, since
graviton correlators and particular metric perturbations can depend on
the gravitational gauge~\cite{Garriga:2007zk,Tsamis:2007is,
  Higuchi:2011vw}.  Kitamoto and Kitazawa have argued that
appropriately defined time-dependent physical couplings and their
relative scaling provide gauge-independent characterizations of
infrared dynamics
\cite{Kitamoto:2012vj,Kitamoto:2014gva,Kitamoto:2016yfy}.  These bulk
calculations characterize the infrared gravitational dynamics, but do
not by themselves provide a microscopic quantum-gravitational origin
for the stochastic source. In semiclassical stochastic gravity
\cite{Hu:2008rga}, the noise kernel is constructed from quantum matter
stress-tensor fluctuations. An intrinsic noise kernel derived from the
microscopic quantum degrees of freedom of spacetime itself would
therefore provide a complementary quantum-gravitational description.

This suggests a more microscopic physical picture. A quantum horizon
should contain fast quantum-geometric degrees of freedom in addition
to the slow collective variables that characterize the macroscopic
geometry. If the fast sector is sufficiently interacting to lose
memory, its force on a slow geometric mode should appear, after coarse
graining, as stochastic gravitational noise.  The Starobinsky
mechanism for gravity would require the quantum geometry itself to
provide the fluctuating microscopic degrees of freedom, without the
need for an external matter environment.
The relevance of quantum chaos to de Sitter horizons has been
demonstrated by Aalsma and Shiu \cite{Aalsma:2020aib}, who found
maximal chaos from out-of-time-order correlators in a gravitational
shockwave background. While their analysis probes chaos through
semiclassical gravitational scattering, a microscopic quantum-horizon
model offers the possibility of studying not only the origin of
chaotic dynamics but also its consequences for force relaxation,
stochastic fluctuations, and the infrared evolution of gravitational
collective modes.

In this paper, we show how this physical picture can be realized
microscopically in the recently proposed matrix model of the
black-hole horizon \cite{Chu:2024qil}.  In this framework the horizon is
represented by a fuzzy two-sphere built from matrix degrees of
freedom, so that the microscopic variables are themselves degrees of
quantum geometry rather than fields propagating on a prescribed
background. The model reproduces the basic macroscopic relations among
the black-hole mass, horizon radius, Bekenstein--Hawking entropy, and
Hawking temperature, while its large set of microscopic horizon states
provides the corresponding statistical interpretation of the entropy
\cite{Chu:2024qil,Chu:2024edh}.  Previous work has further developed a
microscopic description of Hawking emission \cite{Chu:2026wyh},
membrane-like horizon response \cite{Chu:2026vzj,Chu:2026qom}, and the
quantum stability of the fuzzy-sphere background \cite{Chu:2026dhx}.
Its microscopic spectrum exhibits random-matrix universality
characteristic of quantum chaos \cite{Chu:2026rac}. More recently, the
one-loop dynamics of the homogeneous $(1,0)$ fluctuation has revealed
a quantum tadpole \cite{Chu:2026kqt} that couples directly to the
overall scale of the horizon and plays an important role in the
black-hole/de Sitter correspondence developed below: the same $(1,0)$
mode changes the overall scale of the fuzzy sphere and therefore
provides a natural slow collective variable on which the fast
microscopic horizon degrees of freedom can act.

This construction also offers an interesting connection with
Polyakov's conjecture that infrared screening may require
quantum-gravitational physics beyond perturbations around classical de
Sitter space. In our model, the region beyond the de Sitter static
patch is replaced by a Schwarzschild exterior, with the two geometries
joined at a microscopic quantum horizon. This gravitational saddle is
globally distinct from pure de Sitter space, while the chaotic horizon
dynamics provides a possible microscopic source of stochastic
fluctuations and infrared screening.  In this sense, our construction
is intriguingly consistent with Polyakov's broader conjecture, although
the underlying mechanism is different.

The central proposal of the present work is that the homogeneous
matrix breathing mode $q_{10}$, or equivalently the dimensionless
breathing deformation
\be
\varpi=\frac{q_{10}}{\sqrt{D_N}},
\ee
must first be gravitationally dressed before its infrared dynamics is
identified.  In the isolated matrix theory this mode has positive
quadratic curvature, but once the fuzzy-sphere horizon is embedded
into the Schwarzschild--de Sitter junction, $\varpi$ combines with the
relative-lapse fluctuation $\phi_C$. With the background-matched
dictionary introduced in Sec.~3.3, the corresponding gravitational
scalar variable
\be
X=\sqrt{3}\,(\varpi+\phi_C)
\ee
is the massless
minimally coupled scalar mode of de Sitter gravity.  The completed
matrix--junction system possesses a classically massless collective
mode which approaches $X$ at large $N$.  The physical slow variable
relevant to the infrared dynamics is therefore not the isolated
breathing deformation, but its gravitationally dressed collective
coordinate.
The remaining microscopic horizon degrees of freedom then act on this
soft mode through the force conjugate to the breathing coordinate.
The Gaussian theory already contains substantial force fluctuations,
but its discrete spectrum produces coherent dephasing rather than
irreversible stochastic noise.  We argue that interactions and the
associated chaotic many-body dynamics relax the projected force and
generate finite-memory noise and dissipation.  After coarse graining
over the fast horizon sector, the soft mode consequently obeys a
generalized Langevin dynamics and, after momentum relaxation,
undergoes Brownian diffusion.

A central result of the present work is that stochastic diffusion
generates coarse-grained entropy. The same microscopic force
fluctuations that drive the Langevin dynamics also induce decoherence
of the soft mode, allowing its late-time Fokker--Planck distribution
to describe the coarse-grained probability distribution of the reduced
state. Under suitable mixing and central-limit conditions, the
accumulation of weakly correlated horizon kicks produces an
approximately Gaussian distribution, whose spreading leads to
logarithmic growth of the coarse-grained stochastic entropy. Thus, the
entropy growth originates from microscopic horizon dynamics rather
than being imposed from the gravitational description.

Assuming approximate statistical independence, this coarse-grained
entropy can be combined with the background microstate entropy arising
from the degenerate fermionic parton states. We conjecture that the
resulting quantum-corrected thermodynamic entropy can be identified
with the Gibbons--Hawking entropy of the backreacted causal
horizon. This identification allows the entropy growth to be
interpreted as a gravitational response.
Following Kitazawa and collaborators \cite{Kitamoto:2019rij}, we
express this response in terms of the dimensionless gravitational
coupling
\be
g(T):=\frac{G_N H^2(T)}{\pi}.
\ee
We show that it obeys the screening beta function
\be
\frac{dg}{d{\cal L}_T}=-\frac12 g^2.
\ee
Here the flow parameter denotes the logarithmic broadening of the
soft-mode variance and is defined explicitly in Sec.~5. The screening
coefficient agrees with that obtained by Kitazawa and collaborators
when the flow is expressed in terms of logarithmic covariance
broadening. In our construction, this coefficient follows from
Gaussian entropy growth and is independent of the detailed variance
evolution, which is determined by the microscopic noise and
dissipation coefficients.

Since stochastic diffusion increases the variance, the
entropy--gravity identification implies a decreasing gravitational
coupling, corresponding to a reduction of the Hubble parameter and the
effective cosmological constant. This provides a microscopic mechanism
for infrared screening of de Sitter curvature, conditional on the
proposed gravitational interpretation of the thermodynamic
entropy. Establishing the corresponding change in the physical
causal-horizon geometry requires a further analysis of gravitational
backreaction.

This work thus provides a microscopic realization of the physical
picture expected from a fundamental construction of stochastic de
Sitter gravity.  The stochastic force is not postulated
phenomenologically, nor does it originate from matter propagating on a
fixed background.  It is generated by the microscopic quantum degrees
of freedom of the horizon itself.  The resulting picture may be
summarized as
\be
\text{horizon chaos}\ \rightarrow\ \text{force noise}\
\rightarrow\ \text{soft mode diffusion}\ \rightarrow\
\text{entropy growth}\
\rightarrow\ dS\ \text{screening}. \nn
\ee
The $(1,0)$ tadpole
and the stochastic fluctuations of the same microscopic horizon sector
thus play complementary roles: the former motivates the
Schwarzschild--de Sitter gravitational completion, while the latter
drive the infrared stochastic dynamics of its gravitationally dressed
soft mode.

A decisive quantitative test of this picture is to compute the
interacting retarded and symmetrized correlators of the projected
microscopic force.  These correlators determine the low-frequency
noise strength, dissipative response, and memory time entering the
effective stochastic dynamics.  Deriving them directly from the full
interacting many-body horizon theory would turn the present stochastic
construction into a first-principles microscopic prediction.

The remainder of the paper is organized as follows.  In Sec.~2, we
review the microscopic horizon matrix model, the homogeneous $(1,0)$
breathing mode and its tadpole, and the Schwarzschild--de Sitter
junction in which the fuzzy sphere forms the common horizon.  In
Sec.~3, we construct the massless collective mode and match it to the
gravitational scalar modes.  In Sec.~4, we analyze the microscopic
force acting on it, including the force correlators,
interaction-induced low-frequency noise, dissipation, and the
Markovian limit.  In Sec.~5, we derive the resulting
coarse-grained entropy production and de Sitter
screening.
We summarize our main results in Sec.~6 and
discuss broader implications in Sec.~7.  An appendix collects details
of the Gaussian force-spectrum calculation.

\section{Quantum Horizon Geometry and its Microscopic Force}
\subsection{The fuzzy-sphere horizon and the homogeneous scale mode}

The quantum-horizon model is a matrix quantum mechanics with action
\cite{Chu:2024qil} 
\be \label{L}
S = \int \dd t \Tr \left[
   \frac{1}{2a_0^2 M_P} \dot{X}_a^{2}
   + \frac{M_P}{N^2} \left([X_a,X_b]^2 + 4  X_a^{2}\right)
   %c2 typo before
   + i  \psi^\dag \dot{\psi}
   - a_2 \frac{M_P}{N^2}\psi^\dag \s_a X_a  \psi \right].
\ee
The rank-$N$ fuzzy sphere $S^2_N$ background is
\be
 X_a=J_a,
 \qquad [J_a,J_b]=\ii\epsilon_{abc}J_c,
 \qquad J_aJ_a=J(J+1)\bm 1_N,
 \qquad J:=\frac{N-1}{2}.
 \label{fuzzy}
\ee
In \cite{Chu:2024qil}, we found that
the Hessian $\cN_J$ of the fuzzy sphere is
diagonalized by  vector fuzzy harmonics $B_a^{L KQ}$
\be
\cN_J B_a^{LKQ}=\Lambda_{LK}B_a^{LKQ}
\ee
with
\be\label{modes1}
1 \leq L \leq N-1, \qquad K=L-1, L, L+1, \qquad Q =-K, \cdots, K.
\ee
The explicit construction of the fuzzy vector
spherical harmonics $B_a^{LKQ}$ followed
the construction of Thorne \cite{Thorne:1980rmp}
for the continuum vector spherical harmonics,
and  can be found in the appendix A of \cite{Chu:2026dhx}.
We adopt the conventional unit Hilbert--Schmidt normalization
for the  harmonics $B_a^{LKQ}$:
\be
\sum_a\Tr\!\left[(B_a^{L KQ})^\dagger B_a^{L KQ}\right]=1.
\ee
The classical spectrum contains a tachyonic $(L,K)=(1,2)$ mode, with
classical eigenvalue $-2$, and a marginal $(L,K)=(2,3)$ mode, with
classical eigenvalue zero. The $(L,K)=(1,0)$ scaling mode is instead
classically stable, with eigenvalue $4$. It was subsequently shown in
\cite{Chu:2026dhx}
that the tachyonic $(1,2)$ and marginal $(2,3)$ modes receive
positive quantum corrections sufficient to stabilize them and, more
generally, that the complete quantum-corrected fluctuation spectrum is
positive for sufficiently large but finite $N$. The $(1,0)$ mode
receives a negative $1/N$ one-loop correction and remains stable. Together
with the suppression of higher-order quantum corrections at large $N$,
this establishes the quadratic quantum stability of the fuzzy-sphere
background.

The same analysis also revealed that the distinguished $(L,K)=(1,0)$
scaling mode develops a quantum tadpole whose physical effect is a
contractive pressure on the fuzzy sphere. In \cite{Chu:2026kqt},
this observation
was used to argue that the spacetime inside the horizon need not be in
the same vacuum as the exterior Schwarzschild region. A null-junction
analysis showed that, assuming a static, spherically symmetric,
regular vacuum in the interior, the interior geometry is de Sitter
space. The matching further determines a soldering relation $T = C_0 t$
between
the interior de Sitter time $T$ and the exterior Schwarzschild time
$t$, with
a relative lapse scaling as
\be \label{C0}
C_0=\frac{4\pi N}{9b}+O(1) = O(N),
\ee
where $M_P=b^{-1}\sqrt{2/(\pi G)}$ and $l_P=\sqrt{2G/\pi}$
\cite{Chu:2024qil}. Thus, in this
construction, the fuzzy-sphere horizon acts as a quantum-geometric
domain wall separating distinct exterior and interior gravitational
configurations.

The normalized $(L,K)=(1,0)$  vector fuzzy
harmonic is
\be
B_a^{100}=\frac{J_a}{\sqrt{D_N}},
\qquad D_N := \sum_a \Tr J_a^2=\frac{N(N^2-1)}{4}.
\label{eq:B100}
\ee
Here 
the label $L=1$ refers to the matrix-harmonic orbital representation.
The physical angular momentum is obtained by coupling this orbital
representation to the vector index $a$ of $X_a$. For the present mode
the total angular momentum is $K=0$, so despite its orbital label
$L=1$, it is rotational invariant.
In fact, the displayed configuration 
$X_a=J_a+q_{10} B^{100}_a$ along this mode is
\be
X_a=(1+\varpi)J_a,
\qquad
\varpi := \frac{q_{10}}{\sqrt{D_N}}.
\label{eq:Xvarpi}
\ee
Thus the $(L,K)=(1,0)$ mode is a homogeneous scaling, or breathing,
mode of the fuzzy sphere. Geometrically, 
the rescaling in Eq.~\eqref{eq:Xvarpi} gives rise to a rescaling of the
fuzzy sphere radius
\be
R=R_0(1+\varpi).
\label{eq:Rvarpi}
\ee
Therefore
\be
\frac{\delta R}{R_0}=\varpi
\label{eq:dRR}
\ee
at linear order, 
and the area obeys
\be
A=A_0(1+\varpi)^2.
\label{eq:Avarpi}
\ee
We will show below that this mode defines a slow collective coordinate
coupled to the remaining microscopic degrees of freedom of the
horizon, whose fast interacting dynamics, provided it produces
sufficient spectral mixing, induces stochastic fluctuations of the
intrinsic matrix--junction geometry.  We emphasize that the radius
\eq{eq:Rvarpi} and area \eq{eq:Avarpi} defined above refer to
the intrinsic fuzzy-sphere geometry.

The junction construction of \cite{Chu:2026kqt}
identifies the undeformed fuzzy
sphere with the common spherical interface of a Schwarzschild exterior
and a de Sitter interior.  On this static background,
\be
H_0=R_0^{-1}.
\ee
This background relation should not, however, be promoted to a
configuration-by-configuration identification between the stochastic
breathing fluctuation and the de Sitter Hubble parameter.  Once the
collective mode fluctuates, its gravitational interpretation requires
the complete junction dictionary developed in Sec.~3.  In particular,
the intrinsic matrix--junction radius and the reconstructed de Sitter
causal-horizon radius are {\it distinct} observables.

\subsection{The conjugate microscopic force}

The dynamics of the collective coordinate $q_{10}$ is governed by the force
obtained by differentiating the horizon Hamiltonian,
\be
F_{10}=-\frac{\partial H_{\rm hor}}{\partial q_{10}}.
\label{eq:Fdef}
\ee
This can be separated into its expectation value and fluctuation,
\be
F_{10}=\langle F_{10}\rangle+\delta F_{10}.
\label{eq:Fsplit}
\ee
Here the average $\langle F_{10}\rangle$ describes the mean tadpole
force on the collective geometry. For our system,
the mean force has been balanced and thereby
absorbed into the definition of the saddle. Consequently 
the present stochastic analysis will
focus on the connected fluctuation $\delta F_{10}$.

The stochastic fluctuations are captured by the
symmetrized force correlator \cite{Kubo:1966fyg,Grabert:1988yt}
\be
C_{F_{10}}(T)=\frac{1}{2}\left\langle\left\{\delta F_{10}(T),\delta
F_{10}(0)\right\}\right\rangle,
\label{CF}
\ee
whose spectral density is
\be
S_{F_{10}}(\omega)=\int_{-\infty}^{\infty}dT\,e^{i\omega T}C_{F_{10}}(T).
\label{SF}
\ee
By construction,  $C_{F_{10}}(T)$ measures the temporal correlation of
fluctuations of the generalized force acting on the collective coordinate 
$q_{10}$ and the spectrum $S_{F_{10}}(\omega)$
tells us at which frequencies the force fluctuates.
On the other hand, the quantum response is captured by the retarded correlator
\be
\chi_{F_{10}}^R(T)=-i\theta(T)\left\langle\left[\delta F_{10}(T),\delta
  F_{10}(0)\right]\right\rangle,
\label{eq:chiR}
\ee
whose low-frequency behavior determines the dissipative coefficient
\be
\eta_{F_{10}}
:=
-\lim_{\omega\to0}
\frac{\operatorname{Im}\chi_{F_{10}}^R(\omega)}{\omega}.
\label{eta}
\ee

The separation between Eqs.~\eqref{SF} and \eqref{eta} will be useful
when we discuss state dependence. In a thermal KMS state, the
symmetrized noise spectrum and the dissipative part of the retarded
response are related by the fluctuation--dissipation relation. In our
soldered geometry, however, an Unruh-like state may be physically more
appropriate, in which case the total noise and response need not
satisfy a single equilibrium KMS relation. See \cite{Aalsma:2019rpt}
for a closely related discussion.

\section{Junction Completion and the Massless Breathing Mode}
\label{sec:gravmixing}

The homogeneous fuzzy-sphere scaling mode $q_{10}$ has so far been
treated as a collective coordinate of the microscopic horizon
system. The isolated matrix calculation assigns this mode a
nonvanishing tadpole together with a small positive restoring
curvature, showing that the fuzzy sphere is locally stable but is not
by itself a stationary configuration. In Ref.~\cite{Chu:2026kqt},
we proposed
that, in the gravitational completion, the tadpole is balanced when
the fuzzy sphere is identified with the common Schwarzschild--de
Sitter horizon. Here we describe this mechanism more explicitly in the
matrix-model language.
We introduce, in addition to the fractional breathing deformation
$\varpi$, an effective scalar $\phi_C$ associated with fluctuations of
the relative lapse across the junction. We propose a minimal
two-scalar potential as an effective counterpart of the gravitational
dynamics. This makes it possible to trace both the cancellation of the
matrix tadpole and the softening of the breathing mode to the
properties of the same set of scalar degree of freedom. We then
determine the resulting collective modes and match them geometrically
to the scalar eigenmodes of de Sitter gravity.

\subsection{Junction completion of the matrix breathing mode}

Let $\bar{\sigma}_{AB}=R^2\gamma_{AB}$ denote the induced metric on the
unperturbed spherical horizon, where $\gamma_{AB}$ is the unit two-sphere
metric. Under the homogeneous breathing deformation \eq{eq:Xvarpi},
we have
\be\label{dsigma}
\frac{\delta\sigma_{AB}}{\bar{\sigma}_{AB}}=2\varpi, 
\ee
with the ratio understood component-wise.
The isolated matrix effective action contains a positive quadratic
curvature
along this coordinate. In the de Sitter clock, the corresponding bare
frequency is of order
\be
\omega_{10}^{\rm bare}\sim\frac{H}{N}.
\ee
This frequency characterizes the response of the breathing coordinate
$q_{10}$ in the isolated matrix model. Once the horizon system is
coupled to gravity, however, the microscopic breathing coordinate need
not remain a normal coordinate of the homogeneous scalar sector of the
coupled horizon--gravity system. The intrinsic breathing deformation
can mix with additional scalar gravitational data associated with the
temporal metric and the null junction. The normal modes and their
eigen-frequencies must therefore be determined from the resulting
coupled homogeneous scalar fluctuation problem.

The necessity of including the gravitational background was already
the central lesson of the tadpole analysis
\cite{Chu:2026kqt}. There, the isolated matrix theory was found to
generate at one loop a nonzero scale-mode tadpole, whose cancellation
requires an adjustment of the gravitational background according to
\be
\delta\Gamma_{\rm FS}+\delta\Gamma_{\rm grav}=0.
\ee
The gravitational background therefore participates essentially in
defining the physical saddle. Here we apply the same logic at
quadratic order: the curvature of the physical homogeneous scalar mode
can only be determined after the matrix breathing mode $q_{10}$ is
gravitationally completed. 

The additional scalar data required for this completion can be seen
directly from the null-junction construction. The matrix breathing
mode determines the intrinsic angular geometry through
Eq.~\eqref{dsigma}. Once the horizon is embedded gravitationally,
however, the junction contains additional geometric information that
is not encoded in the intrinsic two-metric. On the de Sitter side the
static metric, written using the common exterior and matrix time,
takes the form
\be
ds_-^2=-C^2h_-(r)dt^2+\frac{dr^2}{h_-(r)}+r^2d\Omega_2^2 .
\label{eq:dSmetricgravmix}
\ee
The background value \eq{C0} of $C$ is determined by the
surface-gravity jump together with the microscopic tadpole
balance, its fluctuation gives rise to
an additional scalar junction datum.
Let us parameterize fluctuations of this relative normalization by
\be
C=C_0e^{\phi_C},\qquad \phi_C:=\delta\ln\frac{C}{C_0}.
\ee
The minimal scalar data associated with the homogeneous horizon
deformation are therefore
\be
(\varpi,\phi_C).
\ee
The variable $\varpi$ determines the intrinsic horizon two-geometry,
while $\phi_C$ describes the fluctuation of the relative normalization
of the null generator across the junction. At this stage $\phi_C$ is
introduced only as an effective junction scalar; no microscopic matrix
realization of it is assumed.

We now ask whether the same additional scalar degree of freedom that
participates in balancing the matrix tadpole can also soften the
positive curvature of the isolated breathing mode. In general, the potential
energy of the combined matrix-gravity system  consists of the matrix-model
energy $E_{\rm mat}(\varpi)$, the
gravitational energy 
and a junction interaction term $U_{\rm jun}(\varpi,\phi_C)$.
We decompose the gravitational contribution into the exterior
Schwarzschild energy and the de Sitter energy,
with the latter carrying the relative-lapse normalization $C_0e^{\phi_C}$ when
expressed in the common matrix/exterior clock $t$.
The effective potential is therefore
\be
V(\varpi,\phi_C)=E_{\rm mat}(\varpi)+E_{\rm BH}(R)+C_0e^{\phi_C}E_{\rm dS}(R)
+U_{\rm jun}(\varpi,\phi_C).
\ee
%c10
Here $R=R_0(1+\varpi)$ parametrizes the intrinsic radius of the
junction surface along the static family used to define the effective
potential and its derivatives.
%c12
%% On the undeformed saddle one has
%% \be \label{H0R0}
%% H_0=R_0^{-1}.
%% \ee
%% This static background relation should be distinguished from the
%% identification of general time-dependent fluctuations.  In particular,
%% when the homogeneous collective coordinate is allowed to fluctuate, we
%% shall not assume configuration by configuration that
%% \be
%% H(T)=\frac{1}{R_0[1+\varpi(T)]}.
%% \ee
%% The relation between the matrix--junction variables and the
%% gauge-fixed gravitational scalar fluctuations will instead be specified
%% independently below.  This distinction affects only the dynamical interpretation
%% of the fluctuations. It does not modify the static effective
%% potential, its derivatives with respect to the original field
%% variables, or the tadpole-balance condition.

\subsection{Tadpole balance and the classically massless collective mode}

We now examine the saddle and quadratic fluctuations of the two-scalar effective
potential introduced above. Near the fuzzy-sphere configuration, the
microscopic horizon energy has the expansion
\be
E_{\rm mat}(\varpi)=E_0+\tau\varpi+\frac12K_{\varpi\varpi}^{\rm mat}\varpi^2+\cdots ,
\ee
where the one-loop matrix tadpole  is defined by
\be
\tau:=\left.\frac{\partial E_{\rm mat}(\varpi)}{\partial\varpi}\right|_{\varpi=0}.
\label{taudef}
\ee
It is related to the tadpole surface tension
\be
\tau_{\rm surf}=\frac{a_0M_P}{6\pi\ell_P^2}\sqrt{1-\frac{1}{N^2}},
\ee
introduced in \cite{Chu:2026kqt} by the change of variables
from the horizon area to the fractional radius deformation,
$A(\varpi)=4\pi R_0^2(1+\varpi)^2$. As a result, 
we have
\be
\tau=8\pi R_0^2\tau_{\rm surf}.
\label{tautensionrelation}
\ee
Substituting $R_0=N\ell_P$ and $a_0=\pi/3$, we obtain
at large $N$
\be
\label{taulargeN}
\tau\simeq\frac{4\pi}{9}M_PN^2.
\ee

Stationarity of the completed configuration in the breathing direction
gives
\be
0=\tau+R_0E_{\rm BH}'(R_0)+C_0R_0E_{\rm dS}'(R_0)+\partial_{\varpi}
U_{\rm jun}(0,0).
\label{radialsaddle}
\ee
Stationarity in the relative-lapse direction requires independently
\be
0=C_0E_{\rm dS}(R_0)+\partial_{\phi_C}U_{\rm jun}(0,0).
\label{lapsesaddle}
\ee
Equation~\eqref{radialsaddle} is the direct matrix-model realization
of the tadpole balance proposed in Ref.~\cite{Chu:2026kqt}: the
microscopic force generated by the matrix tadpole is cancelled by the
radial response of the full Schwarzschild--de Sitter gravitational
configuration. Equation~\eqref{lapsesaddle} shows that the junction
contribution $U_{\rm jun}$ cannot in general be omitted.

To illustrate the core idea of the mechanism, we consider a minimal model
and assume that the
residual junction term does not contribute to the leading radial force,
\be
\partial_{\varpi}U_{\rm jun}(0,0)=0.
\label{Uradialzero}
\ee
The radial saddle condition then reads
\be
C_0R_0E_{\rm dS}'(R_0)+R_0E_{\rm BH}'(R_0)=-\tau.
\label{fulltadpolebalance}
\ee
The two gravitational terms on the left-hand side represent the
response of the two sides of the completed geometry under a homogeneous
displacement of the common horizon.
We next expand the potential to quadratic order about this saddle,
\be
V^{(2)}=\frac12K_{\varpi\varpi}\varpi^2+K_{\varpi C}\varpi\phi_C+\frac12
K_{CC}\phi_C^2,
\label{V2}
\ee
where
\be
K_{\varpi\varpi} :=K_{\varpi\varpi}^{\rm mat}+R_0^2E_{\rm BH}''(R_0)
+C_0 R_0^2 E_{\rm dS}''(R_0)+\partial_{\varpi}^2U_{\rm jun}(0,0),
\label{Kwwfull}
\ee
\be
K_{\varpi C}:=
C_0 R_0 E_{\rm dS}'(R_0)+\partial_{\varpi}\partial_{\phi_C}U_{\rm jun}(0,0),
\label{KwCfull}
\ee
\be
K_{CC}:=C_0E_{\rm dS}(R_0)+\partial_{\phi_C}^2U_{\rm jun}(0,0).
\label{KCCfull}
\ee
The Schwarzschild energy does not appear in Eqs.~\eqref{KwCfull}
and \eqref{KCCfull}, because it carries zero relative-lapse weight.

For the minimal realization we further assume that the residual junction
interaction does not contribute to the leading mixed curvature,
\be
\partial_{\varpi}\partial_{\phi_C}U_{\rm jun}(0,0)=0.
\label{Umixedzero}
\ee
Using the radial balance condition \eqref{fulltadpolebalance},
the mixed curvature is then
\be
K_{\varpi C}=-\tau-R_0E_{\rm BH}'(R_0).
\label{KwCexact}
\ee
For the Schwarzschild energy $E_{\rm BH}=R/(2G)$, the second term is of order
$NM_P$, whereas the one-loop matrix tadpole is of order $N^2M_P$. Thus
\be \label{KwClargeN}
K_{\varpi C}=-\tau\left[1+O(N^{-1})\right].
\ee
The exterior contribution is therefore subleading in the mixed curvature. Its
radial curvature in Eq.~\eqref{Kwwfull}, however, must in general be
retained because it can be of the same order as the microscopic
breathing curvature. For the Schwarzschild exterior, $E_{\rm BH}=R/(2G)$ and 
\be
E_{\rm BH}''(R)=0
\ee
vanishes identically.
The same is true of any bulk contribution that is linear in $R$ along
the family of backgrounds being varied.

The exact condition for a classically massless scalar direction is
\be
K_{\varpi\varpi}K_{CC}-K_{\varpi C}^2=0,
\label{detKzero}
\ee
or equivalently
\be \label{Kmassless}
K_{CC} = t_N^2 K_{\varpi\varpi},
\ee
where we have introduced the ratio
\be \label{tN}
t_N := -\frac{K_{\varpi C}}{K_{\varpi\varpi}}.
\ee
To determine the large-$N$ scaling of $t_N$, we use \eq{KwClargeN}
and \eq{taulargeN}.  If the breathing curvature is dominated by the
bosonic matrix contribution $K_{\varpi\varpi}\simeq4M_PN$, we obtain
\be
t_N\simeq\frac{\pi N}{9}.
\ee
More generally,
\be \label{tNscaling}
t_N=O(N),
\ee
provided the gravitational contributions do not change the scaling
$K_{\varpi\varpi}=O(NM_P)$.
The masslessness condition then requires
\be
K_{CC}\simeq t_N\tau=O(M_PN^3).
\ee
Thus the flat direction requires a parametrically large relative-lapse
curvature in the common matrix/exterior-time normalization. This is an
additional condition on the junction dynamics, rather than a
consequence of tadpole balance alone.

In this notation, the quadratic potential reads
\be
V^{(2)}=\frac12K_{\varpi\varpi} H_{\rm res}^2,
\label{VH2}
\ee
where
\be
H_{\rm res}:=\varpi-t_N\phi_C
\ee
is the restoring coordinate selected by the potential.
The condition
\be \label{mdirection}
\phi_C=\frac{\varpi}{t_N}
\ee
specifies the flat direction of the static potential. 
We denote the collective mode tangent to this flat direction by $S$.
In the $(\varpi,\phi_C)$ basis, its
field-space direction is
\be
v_S\propto(t_N,1).
\ee
We note that while the quadratic potential \eq{VH2} depends only on the
restoring combination $\varpi-t_N\phi_C$,
the corresponding physical massive normal mode
is determined by the full quadratic action and therefore also depends
on the kinetic matrix. Its eigenvector will be obtained from the
generalized eigenvalue problem below.
Note also that
the relative-lapse displacement is parametrically
suppressed along the soft direction,
\be
\frac{\phi_C}{\varpi}=\frac{1}{t_N}=O(N^{-1}).
\label{phivarpiscaling}
\ee

The two-scalar model thus realizes the central physical idea in a
single framework: the gravitational completion balances the one-loop
matrix tadpole, while fluctuations of the relative-lapse scalar can
simultaneously remove the positive curvature of one collective
homogeneous mode when the Hessian satisfies the masslessness
condition. We next determine the gravitational content of this soft
mode.

%c10

\subsection{Matching to the gravitational scalar eigenmodes and choice
  of dictionary}

The variables $(\varpi,\phi_C)$ characterize two independent pieces of
the homogeneous matrix--junction configuration.  The matrix breathing
coordinate $\varpi$ determines the intrinsic two-geometry supplied by
the fuzzy sphere, while $\phi_C$ parametrizes the relative normalization
of the null generator across the junction.
The matrix model determines the intrinsic horizon geometry but
does not, by itself, determine the perturbed four-dimensional metric
or the corresponding causal horizon.
A gravitational dictionary must therefore specify how
these intrinsic junction variables are represented in the bulk
perturbation theory.

To make the gravitational variables explicit, we follow the
parameterization of Kitazawa et al. The de Sitter background is written
in conformally flat coordinates as
\be
\bar g_{\mu\nu}=a^2(\eta)\eta_{\mu\nu},\qquad a(\eta)=-\frac{1}{H\eta}.
\ee
Metric fluctuations decompose into an overall conformal mode and
a unimodular part according to
\be
g_{\mu\nu}=\Omega^2\widetilde g_{\mu\nu},\qquad \Omega=
a(\eta)e^{\kappa w},
\qquad \det\widetilde g_{\mu\nu}=-1,
\ee
with
\be
\widetilde g_{\mu\nu}=\left(e^{\kappa h}\right)_{\mu\nu},\qquad
h^\mu{}_\mu=0,\qquad \kappa^2=16\pi G.
\ee
For notational convenience we absorb the factor of $\kappa$ into the
linearized fluctuation fields and define
\be
\omega:=\kappa w,\qquad h^\mu{}_\nu:=\kappa h^\mu{}_\nu.
\ee
Thus $\omega$ is the dimensionless homogeneous conformal fluctuation of
the de Sitter metric. At linear order,
\be
g_{\mu\nu}=a^2(\eta)\left[\eta_{\mu\nu}+2\omega\,\eta_{\mu\nu}+h_{\mu\nu}
  +\cdots\right].
\ee
For the homogeneous scalar sector, tracelessness fixes the isotropic
spatial component in terms of $h^{00}$. Writing the scalar perturbation as
\be\label{scalar-pert}
ds^2=a^2(\eta)\left[-(1+2\alpha)d\eta^2+(1+2\beta)d{\bf x}^2\right],
\ee
one finds
\be\label{alpha-beta}
\alpha=\omega-\frac12h^{00}, \qquad
\beta=\omega+\frac16h^{00}.
\ee
The fractional perturbation of the intrinsic metric of a spherical section
at fixed background coordinates is therefore
\be
\frac{\delta\sigma_{AB}}{\bar\sigma_{AB}}=2\beta=2\omega+\frac13h^{00}.
\ee
Comparing with the matrix breathing deformation \eq{dsigma}
gives
\be \label{dd1}
\varpi=\omega+\frac16h^{00}.
\ee
The second independent junction datum is the fluctuation of the relative
normalization of the null generator. Matching this relative-lapse
perturbation to the homogeneous temporal metric fluctuation gives
\be\label{dd2}
\phi_C=\omega-\frac12h^{00}.
\ee

We define this dictionary perturbatively using the common undeformed
background where the reference configuration
\be
\varpi=0,\qquad \phi_C=0
\ee
is identified with the unperturbed de Sitter configuration
\be
\omega=0,\qquad h^{00}=0.
\ee
We then match the intrinsic angular geometry and the relative temporal
normalization of the junction to the corresponding homogeneous
gauge-fixed metric perturbations used in the de Sitter calculation.
We remark that this prescription matches perturbative field variables
about the common saddle;  it does not, by itself,
reconstruct the gauge-invariant location of the perturbed causal horizon.

At quadratic order, the homogeneous scalar sector of de Sitter gravity
is diagonalized by the combinations introduced by Kitazawa et al.,
\be
X:=2\sqrt{3}\,\omega-\frac{1}{\sqrt{3}}h^{00},
\qquad
Y:=h^{00}-2\omega,
\ee
where  $X$ is the massless minimally coupled scalar direction,
whereas $Y$ is the conformally coupled scalar direction with a
frequency of order the de Sitter scale.  The dictionary \eq{dd1}, \eq{dd2}
therefore
gives 
\be\label{XYvarpi}
X=\sqrt{3}\,(\varpi+\phi_C),
\qquad
Y=-2\phi_C.
\ee
Equivalently,
\be\label{varpiXY}
\varpi=\frac{X}{\sqrt{3}}+\frac{Y}{2},
\qquad
\phi_C=-\frac{Y}{2}.
\ee

We distinguish this field dictionary from  another meaningful
construction.  One may instead reconstruct the actual
causal horizon of the perturbed four-dimensional metric.  For a
homogeneous scalar perturbation \eq{scalar-pert},
the two metric perturbations \eq{alpha-beta} are
\be
\alpha = -\frac{Y}{2}, \qquad \beta =  \frac{X}{\sqrt{3}}+\frac{Y}{2}.
\ee
For constant homogeneous perturbations, the radial null condition
implies a displacement of the coordinate position of the causal
horizon,
\be
\frac{\delta x_H}{\bar x_H} = \alpha-\beta.
\ee
The corresponding perturbation of its physical areal radius is then
\be \label{RY2}
\frac{\delta R_H^{\rm causal}}{\bar R_H} = \beta+\frac{\delta
  x_H}{\bar x_H} = \alpha = -\frac{Y}{2}.
\ee
%c10
%% Thus the intrinsic metric perturbation on the background-matched junction
%% and the radius of the reconstructed perturbed causal horizon are
%% distinct observables.  If instead the matrix breathing coordinate were
%% identified directly with the reconstructed causal-horizon radius, one
%% would obtain
%% \be
%% \varpi=-\frac{Y}{2}.
%% \ee
%% In that alternative matching, the relative-lapse fluctuation would not
%% be identified algebraically with the minimally coupled field $X$.
%% Rather, it determines the rate of the $X$ dilation relative to the
%% junction clock,
%% \be
%% \dot X=\sqrt{3}\,H\,(\phi_C-\varpi).
%% \ee
%% We do not adopt this alternative identification here.
Thus, for constant homogeneous perturbations, the intrinsic junction
geometry and the reconstructed causal horizon are distinct
observables. Identifying the matrix breathing coordinate with the
latter would instead give
\be \label{varpiY2}
\varpi=-\frac{Y}{2}.
\ee
For time-dependent perturbations, the causal-horizon
must be reconstructed from the null-horizon condition.
We adopt the
background-matched dictionary \eq{varpiXY}, which matches the
intrinsic angular metric and relative temporal normalization at the
junction. This choice couples the microscopic force conjugate to
$q_{10}$ directly to the massless minimally coupled scalar $X$,
providing a possible microscopic source of its infrared fluctuations.

%c11
%% Note that the static relation \eq{H0R0}
%% continues to characterize the undeformed Schwarzschild--de Sitter
%% saddle and the static family entering the tadpole potential, but we do
%% not infer from it that an arbitrary stochastic realization satisfies
%% \be
%% H(T)=\frac{1}{R_0[1+\varpi(T)]}.
%% \ee
%% The distinction will become important when the stochastic infrared
%% dynamics is interpreted below.

We may now resolve the collective modes of the completed
matrix--junction system in the independently known gravitational basis.
Along the flat direction \eq{mdirection} of the static potential,
we have
\be\label{YSXS}
\frac{Y_S}{X_S}= -\frac{2}{\sqrt{3}(1+t_N)}.
\ee
Thus, up to an overall normalization,
\be\label{SXYexact}
S = X-\frac{2}{\sqrt{3}(1+t_N)}\,Y.
\ee
The dictionary itself is exact at linear order.  The large-$N$ limit
enters only in the composition of the junction soft mode.  Since
$t_N=O(N)$,
one has ${Y_S}/{X_S}=O(N^{-1})$
and therefore
\be\label{SXlargeN}
S=X+O(N^{-1})Y.
\ee
That the flat direction determined by the effective junction potential
aligns at large $N$ with the independently identified massless
minimally coupled scalar $X$ of de Sitter gravity is a nontrivial
consistency check of both the matrix model and the junction
construction, and provides evidence that the model captures aspects of
the scalar dynamics of de Sitter gravity beyond the static background
geometry.

We remark that the massive matrix--junction mode may similarly
be written as
\be
H=a_HX+b_HY.
\ee
However, unlike the null direction, the nonzero-frequency eigenvector
depends on the kinetic matrix of $(\varpi,\phi_C)$ and cannot be
inferred from the static potential alone.  Determining the kinetic
structure is therefore necessary both to identify the gravitational
composition of the massive mode and to obtain the effective inertia of
the massless collective coordinate. We turn to this next.

\subsection{Kinetic completion and the scalar spectrum}

Using the relation between the common matrix/exterior clock $t$ and the
canonically normalized de Sitter time $dT=C_0 dt$,
the quadratic potential
and the kinetic term transform as
\be
V^{(T)}=\frac{1}{C_0}V^{(t)}, \qquad 
K_{ij}^{(T)}=\frac{1}{C_0}K_{ij}^{(t)} .
\ee
By contrast, the kinetic coefficients transform as
\be
M_{ij}^{(T)}=C_0 M_{ij}^{(t)} .
\ee
The null direction of the static potential is unaffected by this change
of clock, since the overall factor cancels from the Hessian condition and
from the ratio $t_N=- {K_{\varpi C}}/{K_{\varpi\varpi}}$.
Since $C_0=O(N)$, the corresponding de Sitter-time curvatures scale as
\be
K_{\varpi\varpi}^{(T)}=O(M_P), \qquad
K_{CC}^{(T)}=O(M_PN^2).
\ee
The relative hierarchy is invariant under this change of clock,
\be
\frac{K_{CC}^{(T)}}{K_{\varpi\varpi}^{(T)}}=
\frac{K_{CC}^{(t)}}{K_{\varpi\varpi}^{(t)}}=
t_N^2=O(N^2).
\ee
In what follows we suppress the superscript $(T)$ and understand all
coefficients entering the dynamical quadratic action to be expressed
in the canonical de Sitter-time normalization.

We now complete the quadratic two-scalar theory by including the most
general time-reversal invariant kinetic term,
\be
L_{\rm kin}=\frac12M_{\varpi\varpi}\dot{\varpi}^{\,2}+
M_{\varpi C}\dot{\varpi}\dot{\phi}_C+\frac12M_{CC}\dot{\phi}_C^{\,2}.
\label{K2}
\ee
We assume that the kinetic matrix
\be
M :=\begin{pmatrix}M_{\varpi\varpi}&M_{\varpi C}\\M_{\varpi C}&M_{CC}\end{pmatrix}.
\ee
is regular and positive,
\be
M_{\varpi\varpi}>0,\qquad M_{\varpi\varpi}M_{CC}-M_{\varpi C}^2>0.
\ee
We will find below that, as a dynamical degree of
freedom, $\phi_C$ can be
matched up with the second, $O(H)$ gravitational excitation appearing
in the diagonalization of the dS gravity. 
If $\phi_C$ were instead
algebraic, it could be eliminated through its constraint. In this case,
the null soft trajectory would remain, but there would be no independent
propagating massive partner.  

The dynamical normal modes of the quadratic Lagrangian
$L_2 = L_{\rm kin} - V^{(2)}$ with \eq{VH2}, \eq{K2}
obey the generalized eigenvalue equation
\be
K\,v=\Omega^2M\,v.
\label{geneigen}
\ee
Using the parameterization \eq{tN} and
the masslessness condition \eqref{Kmassless}, the static Hessian reads
\be
K=K_{\varpi\varpi}\begin{pmatrix}1&-t_N\\-t_N&t_N^2\end{pmatrix}.
  \ee
A null vector of $K$
\be
K\,v_S=0
\ee
therefore solves the generalized eigenvalue equation with zero frequency,
\be
\Omega_S^2=0.
\label{OmegaSzero}
\ee
Note that this zero eigenvalue is independent of
the kinetic matrix: the kinetic
completion changes the normalization and inertia of the soft mode, but
not the null direction selected by the potential.  The stochastic
microscopic force can therefore be projected onto this direction
without first specifying the complete kinetic matrix.

Using the gravitational dictionary of the preceding subsection, we
may use $X$ as a convenient coordinate along the soft direction.  At
finite $N$ the collective mode contains the correlated $Y$ component
displayed above; $X$ here labels the same one-dimensional soft
trajectory rather than asserting that $Y$ vanishes identically.
Using \eq{mdirection} and \eq{XYvarpi},
we obtain
\be \label{vpX}
\varpi =
\frac{t_N}{\sqrt{3}(1+t_N)}X, \qquad \phi_C =\frac{1}{\sqrt{3}(1+t_N)}X.
\ee
The inertia coefficient \eq{K2}
of the massless gravitational mode becomes,
\be
M_X :=\frac{t_N^2M_{\varpi\varpi}+2t_NM_{\varpi C}+M_{CC}}{3(1+t_N)^2}.
\label{MXfull}
\ee
For the microscopic breathing coordinate \eq{eq:Xvarpi},
its matrix contribution to the kinetic coefficient is
\be
M_{\varpi\varpi}=D_NM_{10}.
\label{Mvarpi}
\ee
It is useful to parameterize the correction from the relative-lapse
kinetic sector by
\be
Z_M:=1+\frac{2M_{\varpi C}}{t_NM_{\varpi\varpi}}+\frac{M_{CC}}{t_N^2M_{\varpi\varpi}}.
\label{ZMdef}
\ee
Equation~\eqref{MXfull} may then be written in terms of $M_{10}$ as
\be
M_X=\frac{D_NM_{10}}{3}\left(\frac{t_N}{1+t_N}\right)^2Z_M.
\label{MXZM}
\ee
For kinetic coefficients that are not parametrically enhanced relative
to the matrix breathing inertia,
\be
Z_M=O(1),
\ee
we have
\be
M_X\sim\frac{D_NM_{10}}{3}
\ee
Thus gravitational mixing can change the inertia by an order-one
factor without changing its large-rank scaling. Note that the junction
dictionary identifies the microscopic soft trajectory with the
gravitational infrared mode and fixes its normalization,
without requiring the two theories to have identical quadratic kinetic
actions. The difference is irrelevant for the late-time Brownian
variance growth that we are interested in,
which is determined by the effective noise and dissipation coefficients.
The inertia instead controls the momentum-relaxation timescale.

With the quadratic theory now expressed in the gravitational basis,
the nonzero eigen-frequency follows directly from the generalized
eigenvalue problem. Since the Hessian has rank one, the second
eigenvalue is
\be
\Omega_H^2=K_{\varpi\varpi}\frac{M_{CC}+2t_NM_{\varpi C}+t_N^2M_{\varpi\varpi}}
      {M_{\varpi\varpi}M_{CC}-M_{\varpi C}^2}.
\label{OmegaHgeneral}
\ee
The corresponding eigenvector is thus given by
\be
v_H\propto\left(M_{CC}+t_NM_{\varpi C},-
\left(M_{\varpi C}+t_NM_{\varpi\varpi}\right)\right).
\label{vHgeneral}
\ee
Writing $v_H=(\varpi_H,\phi_{C,H})$ in the matrix--junction basis
and transforming with Eq.~\eqref{XYvarpi}, we find
\be
X_H\propto\sqrt3\left[M_{CC}-M_{\varpi C}+t_N
  \left(M_{\varpi C}-M_{\varpi\varpi}\right)\right],\qquad Y_H\propto2
\left(M_{\varpi C}+t_NM_{\varpi\varpi}\right).
\ee
The gravitational composition of the massive mode thus depends
explicitly on the kinetic coefficients.

A particularly interesting limit arises when
\be
M_{\varpi C}=M_{\varpi\varpi}+O(N^{-1}M_{\varpi\varpi})
\label{MXYaligned}
\ee
and the remaining kinetic coefficient $M_{CC}$ is of the same parametric
order as $M_{\varpi\varpi}$. We parameterize it as
\be
M_{CC}=c\,M_{\varpi\varpi},\qquad c>1,
\label{cdef}
\ee
where the inequality follows from positivity of the kinetic matrix at
leading order in the aligned limit.

In this case the leading $X$ component of the massive mode cancels, giving
\be
\frac{X_H}{Y_H}=O(N^{-1}),
\ee
and hence, up to normalization,
\be
H=Y+O(N^{-1})X.
\label{HYlargeN}
\ee
Together with Eq.~\eqref{SXlargeN}, the two collective modes then
approach the two gravitational eigenmodes at large $N$,
\be
S=X+O(N^{-1})Y,\qquad H=Y+O(N^{-1})X.
\ee
The condition \eqref{MXYaligned} has a direct gravitational interpretation.
In fact
the off-diagonal kinetic coefficient in the Kitazawa $(X,Y)$ basis is
\be
M_{XY}=\frac{M_{\varpi\varpi}-M_{\varpi C}}{2\sqrt3}.
\ee
Hence the kinetic term is exactly diagonal in this basis if and only if
\be
M_{\varpi C}=M_{\varpi\varpi}.
\label{Mdiagcondition}
\ee
The large-$N$ condition \eqref{MXYaligned} therefore ensures that the
kinetic terms become diagonal in the orthogonal gravitational eigenmode
basis $(X,Y)$, as expected.

We next examine the dynamical scale of the massive excitation. In the
kinetically aligned limit \eqref{Mdiagcondition} and for \eq{cdef},
Equation~\eqref{OmegaHgeneral} then reduces to
\be
\Omega_H^2
=\frac{K_{\varpi\varpi}}{M_{\varpi\varpi}}
\frac{t_N^2+2t_N+c}{c-1}.
\label{OmegaHaligned}
\ee
For kinetic coefficients of the same parametric order, $c, c-1=O(1)$.
Furthermore, the isolated breathing mode has
\be
\frac{K_{\varpi\varpi}}{M_{\varpi\varpi}}
=O\left(\frac{H^2}{N^2}\right),
\label{KoverMscaling}
\ee
while Eq.~\eqref{tNscaling} gives $t_N=O(N)$. It therefore follows that
\be
\Omega_H^2=O(H^2),
\qquad
\Omega_H=O(H).
\label{OmegaHscale}
\ee
Remarkably, the completed matrix--junction theory reproduces the
qualitative gravitational scalar spectrum: an exactly massless mode
approaching the minimally coupled scalar $X$, and a massive mode of
order the de Sitter scale approaching the conformally coupled scalar
$Y$ in the kinetically aligned regime.
%c12 
%% What remains undetermined at this level is
%% the microscopic dynamics of the relative-lapse variable $\phi_C$ and,
%% correspondingly, the precise values of the kinetic and potential
%% coefficients involving that variable.  These coefficients affect the
%% normalization and characteristic time scales, but not the existence of
%% the null direction.

The matrix--junction soft mode is therefore predominantly the
gravitational dilation mode $X$, with a parametrically suppressed $Y$
component.  We now examine how microscopic horizon dynamics drives this
soft mode and interpret its stochastic diffusion in terms of the
resulting coarse-grained entropy and its gravitational consequences.

\section{From Microscopic Noise to Stochastic Soft-Mode Dynamics}
\label{sec:stochastic}

\subsection{Gaussian fluctuations: coherent noise is not stochastic noise}

In \cite{Chu:2026rac}, we showed that strong spectral chaos arises from
the quantum fluctuating geometric ground state of the free theory. It
is therefore natural to ask whether the free fluctuations of the
stable fuzzy-sphere modes are already sufficient to generate
stochastic dynamics for the massless collective mode identified in
the previous section. At large $N$, this massless mode approaches the
massless gravitational scalar $X$ according to
Eq.~\eqref{SXlargeN}. The answer is no. As we will show,
the free theory exhibits both finite-frequency coherent fluctuations
and, for states with occupation-number fluctuations, a singular
zero-frequency contribution. However, neither is sufficient to
generate stochastic evolution of the zero mode. Nevertheless, the
Gaussian analysis reveals the availability of
a large ultraviolet force bandwidth that
becomes important once interactions destroy the coherent Gaussian
spectrum.

Let $Y_n$ denote the canonically normalized stable Gaussian modes,
where $n=(L,K,Q)$ collectively labels the angular quantum numbers. In
the presence of a homogeneous rescaling, the quadratic Hamiltonian in
the de Sitter clock $T$ is
\be
H_2(q_{10})=\frac{1}{2}\sum_n{}'\left[P_n^2+\Omega_T^2 k_n(q_{10})
  Y_n^2\right],\qquad \Omega_T=\frac{\Omega_t}{C_0}, \label{eq:H2}
\ee
where $\Omega_t:=2a_0M_P/N$ is the microscopic frequency scale in the
matrix-theory clock $t$. The prime denotes a sum over the physical
fluctuation modes orthogonal to the collective $(1,0)$ mode, with the
pure-gauge unitary-orbit modes $L=K$ omitted.
The Hessian eigenvalue of the displaced fuzzy sphere is
\be
k_n(q_{10})=\left(1+\frac{q_{10}}{\sqrt{D_N}}\right)^2(k_n+2)-2,
\label{eq:knq}
\ee
where for simplicity $k_n:=\Lambda_{LK}$ and the degeneracy in $Q$ is
understood. Differentiating with respect to the microscopic breathing
coordinate gives
\be
F_{10}^{(2)}=-\left.\frac{\partial H_2(q_{10})}
{\partial q_{10}}\right|_{q_{10}=0}
=-\frac{\Omega_T^2}{\sqrt{D_N}}\sum_n{}'(k_n+2)Y_n^2,
\label{eq:Fgauss}
\ee
where
\be
\Omega_T=\frac{3H}{\pi N}\left[1+O(N^{-1})\right]
\label{eq:OmegaTderived}
\ee
in terms of the de Sitter scale.

Equation~\eqref{eq:Fgauss} gives the microscopic force conjugate to the
matrix breathing coordinate $q_{10}$. The infrared dynamics instead
takes place along the massless direction of the completed
matrix--junction theory. Along this direction, Eqs.~\eqref{vpX}
and \eq{eq:Xvarpi} give
\be
q_{10}=\sqrt{\frac{D_N}{3}}\frac{t_N}{1+t_N}\,X.
\ee
Here $X$ is used as a convenient coordinate along the massless direction;
the corresponding collective mode is $S=X+O(N^{-1})Y$.
Thus the stochastic dynamics derived below concerns the $X$-like soft
collective coordinate selected by the junction theory.
No configuration-by-configuration identification of $X$ with
the local de Sitter curvature radius is assumed.
The force
projected along this direction is therefore
\be
F:=-\frac{\partial H_{\rm hor}}{\partial X}=\sqrt{\frac{D_N}{3}}
\frac{t_N}{1+t_N}\,F_{10}.
\label{eq:FXF10}
\ee
Since $t_N=O(N)$,
\be
F=\sqrt{\frac{D_N}{3}}\,F_{10}\left[1+O(N^{-1})\right].
\label{eq:FXF10largeN}
\ee
Accordingly,
\be
C_F(T)=\frac{D_N}{3}\left(\frac{t_N}{1+t_N}\right)^2C_{F_{10}}(T),
\qquad S_F(\omega)=\frac{D_N}{3}\left(\frac{t_N}{1+t_N}\right)^2S_{F_{10}}
(\omega),
\label{eq:FXnorm}
\ee
which reduce at leading order in the large-$N$ expansion to the
normalization used below. We shall use $F$ for the force projected
onto the massless collective direction, while retaining $F_{10}$ for
the microscopic matrix force. The Gaussian calculations below are
most conveniently performed in the original $q_{10}$ normalization.

We next derive the microscopic symmetrized force correlator.
For each Gaussian oscillator,
\be
Y_n^2=\frac{1}{2\omega_n}\left(a_n^2+a_n^{\dagger 2}+2N_n+1\right),
\qquad N_n=a_n^\dagger a_n. \label{eq:Y2number}
\ee
The first two terms generate pair creation and annihilation at
$\omega=\pm2\omega_n$, whereas fluctuations of $N_n$ produce a
time-independent contribution to the connected force
correlator. Accordingly, for a stationary free state with negligible
intermode correlations,
\be
C_{F_{10}}^{\rm free}(T)=C^{\rm free}_{F_{10}, \rm pair}(T)+C^{\rm free}_{F_{10},\rm occ}(T).
\label{eq:CFfreedecomp}
\ee
The pair contribution is
\be
C^{\rm free}_{F_{10}, \rm pair}(T)=\frac{\Omega_T^2}{2D_N}\sum_n'
\frac{(k_n+2)^2}{k_n}\left[\left\langle N_n^2\right\rangle+\left\langle N_n
  \right\rangle+1\right]\cos(2\omega_{n,T}T),\qquad
\omega_{n,T}:=\Omega_T\sqrt{k_n},
\label{eq:CFpair}
\ee
while the occupation-number fluctuations give
\be
C^{\rm free}_{F_{10},\rm occ}(T)=A_{F_{10}},\qquad A_{F_{10}}:=
\frac{\Omega_T^2}{D_N}\sum_n'\frac{(k_n+2)^2}{k_n}\left\langle(\Delta N_n)^2
\right\rangle.
\label{eq:CFocc}
\ee
Correspondingly the free force spectrum has the decomposition
\be
S_{F_{10}}^{\rm free}(\omega)=S^{\rm free}_{F_{10}, \rm pair}(\omega)+
S^{\rm free}_{F_{10},\rm occ}(\omega),
\label{eq:SFfreedecomp}
\ee
where $S^{\rm free}_{F_{10}, \rm pair}$ consists of discrete lines at
$\omega=\pm2\omega_{n,T}$ and
\be
S^{\rm free}_{F_{10},\rm occ}(\omega)=2\pi A_{F_{10}}\delta(\omega).
\label{SFocc}
\ee

Since the collective mode $S$ has zero eigen-frequency at the
classical quadratic level, its stochastic motion is governed by the
zero-frequency part of the microscopic noise spectrum. At large $N$
this is equivalently the stochastic dynamics of the gravitational
direction $X$, up to $O(N^{-1})$ mixing with $Y$.
The decomposition \eqref{eq:SFfreedecomp}
makes the zero-frequency structure transparent. Since all
retained stable bath modes have nonzero $\omega_{n,T}$,
the pair spectrum has no support at the origin,
\be
S^{\rm free}_{F_{10}, \rm pair}(0)=0. \label{eq:pairzero}
\ee
The occupation contribution requires some care. In the free theory the
individual occupation numbers are conserved, so that this sector
produces a nondecaying contribution to the force correlator and hence
a singular term proportional to $\delta(\omega)$ in the spectrum. This
zero-frequency weight represents a static force component with
infinite memory and should not be identified with stochastic noise. In
what follows, $S_F(0)$ will always denote the finite, smooth
zero-frequency limit of the relaxing force spectrum, after nondecaying
contributions associated with conserved quantities have been
removed. With this convention, the Gaussian theory has
\be
\label{eq:SFregfreezero}
S_F^{\rm free}(0)=0.
\ee
Thus the free horizon theory does not produce stochastic noise for the
massless collective mode. In the vacuum this statement is particularly
immediate: $\langle N_n\rangle=\langle(\Delta N_n)^2\rangle=0$, so
\be
C^{\rm free}_{F_{10},\rm vac}(T)=\frac{\Omega_T^2}{2D_N}\sum_n'
\frac{(k_n+2)^2}{k_n}\cos(2\omega_{n,T}T),\qquad
C^{\rm free}_{F_{10},\rm occ}=0, \label{eq:CFvac}
\ee
and the entire spectrum consists of discrete pair lines away from the
origin.

The absence of regular zero-frequency weight does not mean that the
microscopic force is weak. The equal-time correlator measures the
total force variance, and the mode sum gives
\be
C_{F_{10}, \rm vac}^{\rm free}(0)\sim\frac{H^2}{N}. \label{eq:Feqvar}
\ee
Moreover, the force-weighted rms pair-line frequency satisfies  (see appendix)
\be
\Delta\omega_{F_{10}}=O(H).
\ee
The free force operator therefore probes the full ultraviolet range of
the matrix fluctuation spectrum. The inverse bandwidth defines a
dephasing timescale,
\be
\tau_{\rm band}:=(\Delta\omega_{F_{10}})^{-1}\sim H^{-1}.
\ee
On this timescale the oscillatory force correlator can be strongly
suppressed by phase cancellation among the different spectral lines.
This dephasing is, however, not irreversible relaxation. At finite
$N$ the Gaussian spectrum remains discrete and infinitely sharp, and
its deterministic phase evolution can in principle recur.

Interactions change precisely this aspect of the problem. The
individual Gaussian occupation numbers $N_n$ are no longer conserved,
while the pair-line spectral weight is redistributed among a dense set
of interacting many-body transitions. If chaotic many-body mixing
produces relaxation, coarse graining over frequency scales larger than
the many-body level spacing can then yield a smooth finite-memory
spectral envelope. Since the collective mode $S$ is classically
massless, the relevant question is whether the interacting theory
develops regular spectral weight at zero frequency and, in particular,
a nonzero value of $S_F(0)$.

Before proceeding to the interacting theory, an important
qualification is necessary. Even when interactions are included, part
of the zero-frequency force fluctuation may be protected by an exact
conservation law and therefore does not represent stochastic
low-frequency noise. In particular, the full interacting horizon
Hamiltonian $H_{\rm hor}$ is conserved, so any component of the force
fluctuation $\delta F$ parallel to the total energy cannot relax. We
therefore decompose
\be
\delta F=a_H\,\delta H_{\rm hor}+\delta F^{\perp},\qquad 
a_H:=\frac{\langle\delta F\,\delta H_{\rm hor}\rangle}
{\langle(\delta H_{\rm hor})^2\rangle},
\ee
so that
\be
\langle\delta F^{\perp}\,\delta H_{\rm hor}\rangle=0.
\ee
The component proportional to $\delta H_{\rm hor}$ is nondecaying and
contributes a zero-frequency delta function to the force
spectrum. More generally, if additional exactly conserved quantities
have nonvanishing overlap with $F$, their contributions must likewise
be projected out.

The stochastic spectrum relevant for the massless collective mode is
therefore the regular spectrum of the relaxing force $F^{\perp}$. In
the interacting discussion below, the ETH ansatz and the zero-frequency
force spectrum are to be understood in this projected sense. For
notational simplicity, we shall suppress the superscript $\perp$ where
no confusion can arise. The central dynamical question is then whether
horizon interactions and many-body chaos convert the large but
coherent Gaussian force fluctuations into a finite-memory spectrum
with
\be
S_F(0)\neq0.
\label{eq:SXregtargetgravmix}
\ee
A nonzero regular zero-frequency spectrum is the microscopic noise
required to drive stochastic motion along the massless collective
direction, which approaches the gravitational zero-mode direction $X$
at large $N$.

\subsection{Stochastic noise from interacting spectral mixing}

The Gaussian theory exhibits substantial microscopic force
fluctuations, with the variance and frequency bandwidth determined in
Sec.~4.1. However, its discrete spectral lines produce coherent
dephasing rather than irreversible relaxation. Interactions can
redistribute this spectral weight among the dense many-body
transitions of the horizon theory, providing a possible mechanism for
the emergence of a smooth, finite-memory force spectrum. The
random-matrix universality previously observed in the conditional
horizon-parton Hamiltonians supports the presence of microscopic
spectral complexity, although spectral statistics alone do not
establish force relaxation.

The operator eigenstate thermalization hypothesis (ETH) provides a
useful description of such spectral mixing. For the force component
projected orthogonally to exactly conserved quantities, introduced in
Sec.~4.1, the ETH ansatz takes the form
\be
(F^\perp)_{mn}=\overline{F^\perp}(\bar E)\delta_{mn}+e^{-S(\bar E)/2}
f_F(\bar E,\omega)R_{mn},
\ee
where $\bar E=(E_m+E_n)/2$, $\omega=E_m-E_n$, and $S(\bar E)$ is the
many-body entropy. The smooth function $f_F$ describes the typical
off-diagonal force matrix elements. Their exponential suppression is
compensated by the exponentially large density of accessible states,
allowing the coarse-grained force spectrum to be governed by a smooth
spectral envelope.
However, ETH alone does not guarantee nonzero regular spectral weight
at zero frequency or a finite force-memory time. These properties
depend on the interacting microscopic dynamics. In particular,
interactions generally remove the separate conservation of Gaussian
occupation numbers,
\be
[H_{\rm hor},N_n]\ne 0,
\ee
allowing their nonconserved fluctuations to relax. Components
overlapping with exactly conserved quantities remain nondecaying and
must be projected out, as discussed in Sec.~4.1.

The force spectrum satisfies the sum rule
\be
C_F(0)=\int_{-\infty}^{\infty}\frac{d\omega}{2\pi}S_F(\omega).
\ee
The Gaussian calculation establishes a natural scale for the
microscopic force fluctuations but does not determine how interactions
redistribute their spectral weight near zero frequency. We therefore
assume that the projected interacting force develops a finite-memory
correlator and a smooth, nonvanishing regular spectrum at the
origin. In the microscopic $q_{10}$ normalization, we parameterize the
latter as
\be\label{SF0}
S_{F_{10}}(0)=c_F\frac{H}{N}.
\ee
Here the force spectrum is understood to exclude all nondecaying
contributions associated with exactly conserved quantities. The
dimensionless coefficient $c_F$ depends on the interacting dynamics
and the quantum state of the horizon.
Using the force projection derived in Sec.~4.1, we obtain
\be \label{SF0-x}
S_F(0)=\frac{D_N}{3}\left(\frac{t_N}{1+t_N}\right)^2
S_{F_{10}}(0)=\frac{c_FD_N}{3}\left(\frac{t_N}{1+t_N}\right)^2\frac{H}{N}.
\ee
Since $t_N=O(N)$, this becomes
\be
S_F(0)=\frac{c_FD_NH}{3N}\left[1+O(N^{-1})\right].
\ee

The assumed finite-memory force spectrum \eq{SF0}
provides the microscopic input
for the stochastic dynamics developed below. Establishing its existence
and determining $c_F$ require a direct calculation of the projected
force correlator in the full interacting horizon theory.

\subsection{Large-\texorpdfstring{$N$}{N}
  horizon bandwidth and the Markovian regime}

A local stochastic description requires more than nonzero regular
spectral weight at the origin. The microscopic horizon degrees of
freedom must also lose memory on a timescale short compared with the
timescale over which the massless collective mode changes
appreciably.
At large $N$, the soft collective mode approaches the minimally
coupled scalar direction $X$, which we use as its coordinate below.
We now determine when its reduced dynamics admits a local Markovian
description.

After coarse graining over the fast interacting horizon degrees of
freedom,
the massless mode obeys a generalized Langevin equation of the form
\be
M_X\ddot X(T)+\int_0^T dT'\,\chi_{F,{\rm dyn}}^R(T-T')X(T')=\xi(T),
\label{eq:GLX}
\ee
where the inertia $M_X$ is given by \eq{MXZM}
and $\chi_{F,{\rm dyn}}^R(T)$ is the
retarded response of the projected microscopic force
$\xi(T):=\delta F^\perp(T)$. Its full retarded response is
\be
\chi_F^R(T)=-i\theta(T)\left\langle[\xi(T),\xi(0)]\right\rangle,
\ee
and the dynamical kernel entering Eq.~\eqref{eq:GLX} is obtained by
subtracting its static part,
\be
\chi_{F,{\rm dyn}}^R(\omega):=\chi_F^R(\omega)-\chi_F^R(0).
\ee
The corresponding symmetrized noise kernel is
\be
C_F(T-T')=\frac12\left\langle\{\xi(T),\xi(T')\}\right\rangle.
\ee
All force correlators appearing here are understood to involve the force
projected orthogonally to the exactly conserved quantities.

The subtracted static part contributes to the effective quadratic
curvature of the collective mode. At the classical level this
curvature vanishes because $S$ is the massless collective mode of the
completed horizon--junction system. At large $N$, $S$ approaches the
massless minimally coupled gravitational direction $X$. Any residual
static contribution generated by quantum corrections is included in
the radiatively induced curvature discussed in the next subsection.
By construction,
\be
\chi_{F,{\rm dyn}}^R(0)=0.
\ee
The low-frequency dissipative part of the retarded response may then
be written as
\be
   {\rm Im}\,\chi_{F,{\rm dyn}}^R(\omega)=-\eta_F\omega\,
   [1+O(\omega^2\tau_F^2)],
\ee
where
\be
\eta_F:=-\lim_{\omega\to0}\frac{{\rm Im}\,\chi_{F,{\rm dyn}}^R(\omega)}{\omega}.
\ee
The noise strength and the dissipative coefficient are independent
low-frequency properties of the interacting horizon theory unless an
additional equilibrium condition, such as KMS, is imposed.
We assume a positive dissipative coefficient so that the collective
mode relaxes rather than undergoing an instability.

The noise kernel defines a microscopic correlation time $\tau_F$ such that
\be
C_F(T)\simeq0,\qquad |T|\gg\tau_F.
\ee
Finite $\tau_F$ expresses genuine loss of memory by the projected microscopic
force. For collective dynamics involving frequencies satisfying
\be
|\omega|\tau_F\ll1, \label{eq:MarkovLowFreq}
\ee
the force spectrum may be expanded around the origin,
\be
S_F(\omega)=S_F(0)\, [1+ O(\omega^2\tau_F^2)].
\ee
Inverse Fourier transformation then gives the white-noise approximation
\be
C_F(T-T')\simeq S_F(0)\delta(T-T'). \label{eq:WhiteNoise}
\ee
The retarded kernel similarly reduces to a local friction term, and the
generalized Langevin equation takes the Markovian form
\be \label{eq:LocalZeroMode}
M_X\ddot X(T)+\eta_F\dot X(T)=\xi(T).
\ee
There is no harmonic restoring term because the collective mode is
classically massless.

It is useful to define the momentum-relaxation rate
\be
\label{gammaX}
\gamma_X:=\frac{\eta_F}{M_X}.
\ee
The force projection derived above implies
\be
\eta_F=\frac{D_N}{3}\left(\frac{t_N}{1+t_N}\right)^2\eta_{F_{10}}.
\label{eq:etaFprojection}
\ee
Combining this with Eq.~\eqref{MXZM} gives
\be
\gamma_X=\frac{1}{Z_M}\frac{\eta_{F_{10}}}{M_{10}}.
\label{eq:gammaXexact}
\ee
In the kinetically aligned large-$N$ regime discussed in
Sec.~3.4, $Z_M=1+O(N^{-1})$, and therefore
\be
\gamma_X=\frac{\eta_{F_{10}}}{M_{10}}\left[1+O(N^{-1})\right].
\label{eq:gammaXlargeN}
\ee
Thus the matrix result for the relaxation rate is recovered at leading
order in the large-$N$ expansion.

In the absence of noise the local equation for the zero mode gives
$\dot X(T)\propto e^{-\gamma_XT}$, so that $\gamma_X^{-1}$ is the
characteristic timescale over which the collective momentum changes
appreciably. The Markovian approximation requires the microscopic
force to lose memory on a much shorter timescale,
\be\label{gtauF}
\tau_F\ll\gamma_X^{-1}.
\ee
The condition \eqref{gtauF} quantifies this hierarchy between the
microscopic force-memory time $\tau_F$ and the slower dissipative
timescale $\gamma_X^{-1}$. As found in the Gaussian analysis, the
force spectrum has a characteristic microscopic bandwidth of order
$H$. Although the corresponding Gaussian dephasing is reversible, it
suggests that after interaction-induced spectral mixing, a natural
correlation time
\be
\tau_F=O(H^{-1}) \label{tau-expected}
\ee
of the relaxing force arises. We shall use this as a microscopic
estimate rather than a derived interacting result. Its validity
ultimately requires a direct calculation of the projected interacting
force correlator.

\subsection{Possible radiative lifting and the three stochastic regimes}

The analysis above establishes a vanishing restoring force for the
massless collective mode at the classical quadratic level. At large
$N$, this direction approaches the minimally coupled gravitational
mode $X$, which we continue to use as its coordinate. Since the
masslessness is not known to be protected by a symmetry, higher-order
gravitational or microscopic corrections may generate a small
curvature along this direction. It is therefore useful to consider
the more general Langevin equation
\be
M_X\ddot X+\eta_F\dot X+M_X\Omega_X^2X=\xi, \label{eq:LangevinSmallMass}
\ee
where $\Omega_X^2=0$ at the classical level, while a small nonzero
value may arise from loop corrections. For convenience,
let us parametrize the
radiatively induced curvature with the large $N$ scaling 
\be \label{Omega-p}
\Omega_X^2=b_\n\,\frac{H^2}{N^\n},
\qquad b_\n=O(1),
\ee
where $\n>0$ characterizes the degree of large-$N$ suppression.
This should be regarded as a parametrization of the quantum
lifting of the classically flat direction, rather than as a derived
loop result.

The stochastic description may equivalently be formulated in terms of
the reduced probability distribution of the soft collective degree of
freedom after coarse graining over the fast horizon sector. When
$M_X$ is retained, we introduce the canonical momentum
\be
p:=M_X\dot X,
\ee
and denote the corresponding phase-space probability density by
${\cal P}(X,p,T)$, normalized according to
\be
\int dX\,dp\,{\cal P}(X,p,T)=1.
\ee
The probability density for the collective coordinate alone is its
marginal,
\be
P(X,T):=\int dp\,{\cal P}(X,p,T),\qquad \int dX\,P(X,T)=1.
\ee
In the Markovian regime, the Langevin equation takes the form
\be
\dot p+\gamma_Xp+M_X\Omega_X^2X=\xi, \qquad
\langle\xi(T)\xi(T')\rangle=S_F(0)\delta(T-T').
\ee
Within the Gaussian white-noise approximation, the phase-space
probability density obeys the Kramers equation
\be \label{kramers}
\frac{\partial{\cal P}}{\partial T}
=-\frac{p}{M_X}\frac{\partial{\cal P}}{\partial X}
+\frac{\partial}{\partial p}
\left[\left(\gamma_Xp+M_X\Omega_X^2X\right){\cal P}\right]
+\frac{S_F(0)}{2}\frac{\partial^2{\cal P}}{\partial p^2}.
\ee
At times long compared with the momentum-relaxation time
$\gamma_X^{-1}$, and in the overdamped regime
$\Omega_X\ll\gamma_X$, the momentum can be adiabatically
eliminated. The resulting marginal distribution obeys the
Smoluchowski, or overdamped Fokker--Planck, equation
\be \label{oFP}
\frac{\partial P}{\partial T}
=\kappa_X\frac{\partial}{\partial X}(XP)
+D_X\frac{\partial^2P}{\partial X^2},
\ee
where
\be \label{DXkX}
D_X:=\frac{S_F(0)}{2\eta_F^2},\qquad
\kappa_X:=\frac{\Omega_X^2}{\gamma_X}
=\frac{M_X\Omega_X^2}{\eta_F}.
\ee
For the classically massless mode $\Omega_X^2=0$,
this reduces to the pure diffusion equation
\be \label{FP-diff}
\frac{\partial P}{\partial T}
=D_X\frac{\partial^2P}{\partial X^2}.
\ee

We now distinguish the possible late-time behaviors according to the
sign of the radiatively induced curvature. 
Let us first consider the case 
\be
\Omega_X^2>0.
\ee
In this case we have $\k_X >0$ in  \eq{oFP}.
A parametrically separated three-stage evolution occurs when
\be
\tau_F\ll\gamma_X^{-1}\ll\kappa_X^{-1}. \label{eq:ThreeStageHierarchy}
\ee
Equivalently, this requires both the Markov hierarchy \eq{gtauF}
and the overdamped hierarchy
\be
\Omega_X\ll\gamma_X. \label{eq:SmallMassOverdamped}
\ee
In the first regime,
\be
\tau_F\ll T\ll\gamma_X^{-1},
\ee
the microscopic force has already lost memory, while neither
dissipation nor the weak restoring force has had time to modify the
collective motion appreciably. The mode therefore undergoes stochastic
acceleration. For negligible initial velocity,
\be
\label{eq:StageOneT3}
\left\langle\left[X(T)-X(0)\right]^2\right\rangle\simeq
\frac{S_F(0)}{3M_X^2}T^3.
\ee
In the second regime,
\be
\gamma_X^{-1}\ll T\ll\kappa_X^{-1},
\ee
the momentum has relaxed but the weak restoring force is still negligible.
The dynamics reduces to
\be
\eta_F\dot X=\xi,
\ee
and the mode undergoes ordinary Brownian diffusion,
\be
\label{eq:DXsmallmass}
\left\langle\left[X(T)-X(0)\right]^2\right\rangle\simeq2D_XT.
\ee
At still later times,
\be
T\gtrsim\kappa_X^{-1},
\ee 
the small restoring force becomes important and the overdamped dynamics
reads
\be
\dot X+\kappa_XX=\frac{\xi}{\eta_F}.
\ee
The probability distribution then obeys \eq{oFP}.
It is easy to see that its variance evolves as
\be \frac{d\sigma_X^2}{dT}=2D_X-2\kappa_X\sigma_X^2,
\ee
whose solution is
\be
\sigma_X^2(T)=\frac{D_X}{\kappa_X}
+\left(\sigma_{X,0}^2-\frac{D_X}{\kappa_X}\right)e^{-2\kappa_X(T-T_0)}.
\ee
The variance thus approaches the stationary value
\be \label{sigma-eq}
\sigma_{X,{\rm stat}}^2=\frac{D_X}{\kappa_X}=
\frac{S_F(0)}{2\eta_FM_X\Omega_X^2}
\ee
at late times.
The duration of the diffusive regime is controlled by the inverse
induced drift rate
\be
\kappa_X^{-1}=\frac{\gamma_X}{\Omega_X^2}.
\ee
Thus even a nonzero radiatively generated curvature allows a long
secular diffusion regime provided that the massless direction remains
sufficiently light.

To determine when this saturation becomes important, we now estimate
the relaxation scale associated with the radiatively generated mass.
Using the radiative scaling \eq{Omega-p} and taking a
KMS state as a benchmark.
\be
S_F(0) =  2T_{\rm dS}\eta_F, \qquad T_{\rm dS}=\frac{H}{2\pi}.
\ee
Using
\be
M_{10} = \frac{C_0}{a_0^2 M_P} = \frac{2}{H}[1+ O(N^{-1})],
\ee
together with \eq{SF0-x}, \eq{MXZM},
we obtain the KMS estimate
\be \gamma_X = \frac{\pi c_F H}{2N Z_M} [1+ O(N^{-1})],
\ee
and the restoring drift rate  scales as
\be \kappa_X=\frac{\Omega_X^2}{\gamma_X}\sim HN^{1-\n}.
\ee
Since momentum relaxation occurs at $ \gamma_X^{-1}\sim {N}/{H}$, 
a parametrically extended entropy-producing Brownian regime requires
$\gamma_X^{-1}\ll\kappa_X^{-1}$
or equivalently
\be \n>2.
\ee
The case $\n=2$ remains marginal, while for $\n>2$ the
extended Brownian diffusion regime
persists parametrically beyond momentum
relaxation before eventually saturating.
In the exactly massless limit $ \Omega_X^2=0$,
the cutoff is absent and the linear stochastic
theory predicts continued diffusion. Whether this growth is eventually
bounded depends on the physical configuration space and on effects
beyond the quadratic Markovian approximation, including nonlinear
horizon interactions, finite-size effects, and gravitational
backreaction. The finite fermionic microstate entropy alone does not
determine this cutoff.

If instead higher-order corrections generate a negative curvature,
\be \Omega_X^2<0, \ee
the late-time regime is qualitatively different: the restoring drift
is replaced by an instability, and stochastic fluctuations seed
exponential departure from the quadratic saddle rather than an
Ornstein--Uhlenbeck equilibrium.  Thus the sign and magnitude of the
radiatively generated curvature determine the ultimate fate of the
stochastic flow.   As we show in the next section, they also control
the duration and eventual fate of the
{\it coarse-grained entropy production}
generated by the chaotic horizon dynamics.

\section{Coarse-Grained Entropy Production and de Sitter Screening}

The preceding analysis showed how interaction-induced spectral mixing
generates  finite-memory fluctuations of the microscopic force acting
on the soft collective mode.
At large $N$, this mode approaches the minimally
coupled gravitational scalar $X$. We now study the coarse-grained
entropy generated by its stochastic dynamics and show that it provides
a microscopic realization of the entropy-driven de Sitter screening
mechanism proposed by Kitazawa and collaborators.

\subsection{Microscopic coarse-grained entropy production}

Within the momentum-relaxed Brownian regime, the variance
\eq{eq:DXsmallmass} grows linearly,
\be \label{s2T}
\sigma_X^2(T):=\left\langle [X-\langle X\rangle]^2\right\rangle
=\sigma_X^2(T_0)+2D_X(T-T_0),
\ee
where $T_0$ denotes a reference time in the Brownian regime and $D_X$
is the diffusion coefficient of the soft mode.  The same diffusion may
also be expressed in terms of the intrinsic matrix breathing coordinate
$\varpi$.  Using \eq{vpX}, \eq{DXkX}, 
\eq{SF0-x} and \eq{eq:etaFprojection},
we obtain
\be
D_\varpi=\frac{S_{F10}(0)}{2\eta_{F10}^2D_N},
\ee
with all junction-mixing factors canceling from the breathing-mode
diffusion coefficient.  Thus the stochastic spreading of the soft
gravitational mode is determined directly by the low-frequency noise
and dissipative response of the microscopic horizon theory.

The stochastic dynamics  has a direct open-system
interpretation \cite{Hu:1991di,Grabert:1988yt}.
The soft collective coordinate
couples to the fast
horizon sector through its conjugate microscopic force,
\be
H_{\rm int}=-XF.
\ee
Tracing over the fast horizon degrees of freedom generates a
Schwinger--Keldysh influence functional \cite{Grabert:1988yt}.
To quadratic order in the
coupling, its noise contribution suppresses the separation between the
forward and backward histories according to
\be
\left|{\cal F}[X_+,X_-]\right|=\exp\left[-\frac12\int dT\,dT'\,
  \Delta X(T)C_F(T-T')
  \Delta X(T')\right],\qquad \Delta X:=X_+-X_-,
\ee
where $X_+$ and $X_-$ denote the histories of the soft collective
coordinate on the
forward and backward branches of the Schwinger--Keldysh contour.
%c12
Here we display only the magnitude of the influence functional, which
governs decoherence; its phase contains the retarded response
responsible for dissipation.
In the Markovian regime \eq{eq:WhiteNoise}, 
\be
\left|{\cal F}[X_+,X_-]\right|=\exp\left[-\frac{S_F(0)}{2}\int dT\,
  \Delta X^2(T)\right].
\ee
The difference field $\Delta X$ measures the separation between the
bra and ket histories and therefore vanishes on the diagonal of the
reduced density matrix.
Suppression of configurations with $\Delta X\neq0$ corresponds to
decoherence
in the $X$ basis.
Thus the same microscopic force fluctuations that generate the
Langevin noise also suppress coherence between histories with
different values of $X$.

In the resulting semiclassical regime, the
stochastic probability density is identified with the diagonal part of
the reduced state,
\be P(X,T)=\langle X|\rho_X(T)|X\rangle,
\ee
where
\be
\rho_X(T)=    {\rm Tr}_{\rm fast}\rho_{\rm hor}(T)
\ee
is the reduced density matrix of the soft collective mode obtained
by tracing over the fast horizon degrees of freedom.
The Fokker--Planck equation therefore governs the coarse-grained
probability distribution of the quantum soft mode. This probability
distribution provides a natural measure of the information lost under
the coarse graining over the fast horizon degrees of freedom.  We
define the corresponding coarse-grained stochastic entropy by
\be
S_X^{\rm cg}(T):=-\int dX\,P(X,T)\ln P(X,T).
\ee
Note that under a constant rescaling $X'=aX$, the entropy transforms as
\be
S_{X'}^{\rm cg}(T)=S_X^{\rm cg}(T)+\ln|a|,
\ee
and hence entropy differences and the entropy-production rate are
independent of this additive convention.  This is directly relevant
for the discussion below, since the screening mechanism depends on the
entropy generated relative to a reference state, rather than on the
absolute differential entropy.  In particular, the result is unchanged
if the stochastic dynamics is parametrized by the gravitational soft
coordinate $X$ or by the proportional intrinsic breathing coordinate
$\varpi$. We may therefore evaluate the entropy production directly in the
$X$ representation using the stochastic evolution derived above.  
For the pure diffusion equation \eq{FP-diff}, we have
\be
\frac{dS_X^{\rm cg}}{dT}=D_X\int dX\,\frac{\left(\partial_XP\right)^2}
     {P}\geq0.
\ee
The coarse-grained entropy therefore increases monotonically in the
Brownian regime as a direct consequence of the microscopic horizon
noise.

After many horizon kicks separated by times longer than the
microscopic correlation time, the accumulated displacement is governed
by the central-limit theorem for weakly correlated mixing stochastic
processes \cite{IbragimovLinnik1971}.  Provided that the
force has finite variance and sufficiently rapidly decaying
correlations, the probability distribution therefore approaches a
Gaussian distribution at late times
\be
P(X,T)=\frac{1}{\sqrt{2\pi\sigma_X^2(T)}}
\exp\left[-\frac{[X-\langle X\rangle]^2}{2\sigma_X^2(T)}\right].
\ee
As a result, its coarse-grained entropy
\be
S_X^{\rm cg}(T)=\frac12\ln\left[2\pi e\,\sigma_X^2(T)\right]
\ee
changes with respect to  a reference time $T_0$ according to 
\be \label{DS-univ}
\Delta S_X^{\rm cg}(T):=S_X^{\rm cg}(T)-S_X^{\rm cg}(T_0)
=\frac12\ln\frac{\sigma_X^2(T)}{\sigma_X^2(T_0)}.
\ee
Using the Brownian variance \eq{s2T}, this becomes
\be \Delta S_X^{\rm cg}(T)
=\frac12\ln\left[1+\frac{2D_X(T-T_0)}{\sigma_X^2(T_0)}\right].
\ee
At late times within the Brownian window,
\be \label{DS-lnT}
\Delta S_X^{\rm cg}(T)=\frac12\ln T+{\rm const}.
\ee
The essential result is the universal entropy--variance relation
\eq{DS-univ}, rather than the particular time dependence of the
variance. Microscopic horizon dynamics determines the variance through
its noise and dissipation, while Gaussian statistics fix the entropy
growth per unit logarithmic variance to one-half. This coefficient is
independent of the diffusion constant, relaxation rate, and detailed
covariance evolution. Higher-order corrections may modify these
dynamical quantities but leave the universal coefficient unchanged,
provided the Gaussian stochastic description remains valid.

For a symmetric distribution of the breathing mode, the intrinsic
matrix--junction area satisfies
\be
\frac{\langle A_{\rm jun}\rangle}{A_0}=1+\langle\varpi^2\rangle.
\ee
Its growth reflects the stochastic spreading of the intrinsic junction
geometry, but does not imply growth of the de Sitter causal-horizon
area. As shown by \eq{RY2} and \eq{varpiY2}, the two are distinct even
for constant homogeneous perturbations. For time-dependent
fluctuations, the causal horizon must be reconstructed from the
null-horizon condition. The screening flow derived below instead
follows from the coarse-grained entropy of the stochastic soft mode
and the proposed entropy--gravity identification.

\subsection{Gravitational response and screening of de Sitter curvature}
 
We now consider how the bosonic coarse-grained entropy combines with
the microscopic entropy of the fermionic parton states to define the
total thermodynamic entropy of the horizon.

The Gibbons--Hawking entropy associated with the de Sitter causal horizon
is defined by
\be
S_{\rm GH}:=\frac{A_{\rm causal}}{4G_N}.
\ee
For the undeformed Schwarzschild--de Sitter junction, the causal-horizon
area coincides with the intrinsic junction area $A_0$.
The fuzzy-sphere horizon possesses a highly degenerate fermionic
ground-state sector arising from half-filled parton
configurations. Its leading background microstate entropy is
\be
S_0=S^{(0)}_{\rm GH}=\frac{A_0}{4G_N}=2N^2.
\ee
This reproduces the Gibbons--Hawking area law. The bosonic matrix
coordinates also contribute statistical uncertainty when their
configurations and quantum coherences are unresolved. We denote their
combined coarse-grained entropy by $S_{\rm cg}^{\rm all}$. Its
evolution reflects changes in the bosonic probability distributions,
including the stochastic spreading induced by microscopic horizon
dynamics.

We separate the bosonic degrees of freedom into the gravitationally
dressed soft mode and the remaining fast modes, which act as its
microscopic environment. Assuming that the fast-sector entropy is
approximately stationary and changes in intersector correlations are
negligible, the bosonic entropy change is dominated by the soft mode:
\be
\Delta S_{\rm cg}^{\rm all}(T)=\Delta S_X^{\rm cg}(T).
\ee
In the Brownian regime, Gaussian statistics determine this entropy
change through the variance according to \eq{DS-univ}.
The fermionic microstate entropy and bosonic coarse-grained entropy
describe distinct statistical contributions. Assuming approximate
additivity and negligible changes in the fermionic entropy, we define
the {\it total thermodynamic entropy}
\be \label{S_tot}
S_{\rm th}(T)=S_0+\Delta S_{\rm cg}^{\rm all}(T).
\ee
This thermodynamic interpretation assumes that the coarse graining
captures the relevant unresolved degrees of freedom. The bosonic
contribution then represents a quantum correction to the background
entropy, generated by microscopic horizon dynamics.

Since the background entropy $S_0$ reproduces the Gibbons--Hawking
entropy, we conjecture that the corrected thermodynamic entropy also
admits a gravitational interpretation:
\be
S_{\rm th}(T)=S_{\rm GH}(T).
\ee
Assuming soft-mode dominance, this implies
\be \label{SSA}
\delta S_X^{\rm cg}=\delta S_{\rm GH}.
\ee
The stochastic dynamics of the matrix--junction soft mode does not
directly determine the causal-horizon area. Instead, it generates
coarse-grained entropy, which we conjecture to be accompanied by a
gravitational response through the thermodynamic identification above.
The corresponding change in the causal-horizon area would require
gravitational backreaction, whose dynamical derivation remains an
open problem.

%c12
%% This entropy--gravity identification is our central gravitational conjecture.
%% Its dynamical realization through gravitational backreaction
%% remains to be established.

To examine its consequences, following Kitazawa and collaborators,
we introduce the dimensionless gravitational coupling
\be
g(T):=\frac{1}{S_{\rm GH}(T)}=\frac{G_NH^2(T)}{\pi},
\ee
assuming  a slowly evolving, approximately de Sitter geometry.
For the undeformed background,
\be
g_0^{-1}=S_0=2N^2.
\ee
The Gaussian entropy--variance relation then gives
\be
\frac{1}{g(T)}=\frac{1}{g_0}+\frac12\mathcal L_T,
\qquad
\mathcal L_T:=\ln\frac{\sigma_X^2(T)}{\sigma_X^2(T_0)}.
\ee
Consequently,
\be \label{beta-L2}
\frac{dg}{d\mathcal L_T}=-\frac12g^2.
\ee
As stochastic diffusion broadens the soft-mode distribution, the
gravitational coupling decreases. At fixed Newton's constant, this
corresponds to screening of the effective cosmological constant.

The coefficient $-1/2$ agrees with the result of Kitazawa and
collaborators, although its origin here is different. In our model,
microscopic horizon dynamics generates stochastic fluctuations, while
Gaussian statistics fix the entropy-growth coefficient independently
of the detailed covariance evolution. The agreement concerns this
coefficient, not the time dependence of the gravitational coupling,
which depends on the microscopic dynamics. Higher-order corrections
may modify the variance evolution without changing the coefficient,
provided Gaussian entropy scaling persists.

Finally, the entropy growth need not continue indefinitely. A possible
mechanism for saturation is the radiative lifting of the soft mode
discussed in Sec.~4.4.
%c12 A positive induced curvature produces an Ornstein--Uhlenbeck process
% whose variance approaches the stationary value \eq{sigma-eq}.
For an initially narrower Gaussian distribution,
the entropy increase saturates at
\be
\Delta S^{\rm cg}_{X,{\rm sat}}
=\frac12\ln\left(\frac{D_X}{\kappa_X\sigma_{X,0}^2}\right).
\ee
Radiative lifting therefore limits both the stochastic entropy growth
and, under the proposed entropy--gravity identification, the resulting
de Sitter screening.

\section{Summary and Conclusions}

The main results of this work may be summarized as follows.

\begin{enumerate}

\item {\it A gravitationally dressed soft mode emerges from the
completed matrix--junction system.}

The intrinsic matrix breathing coordinate is not itself a
gravitational zero mode. Once the Schwarzschild--de Sitter junction
is included, the breathing fluctuation mixes with the relative-lapse
degree of freedom. When the masslessness condition on the completed
scalar potential is satisfied, the coupled system possesses a
classically massless collective mode together with a massive partner.
In the gravitational scalar basis, the soft direction satisfies
\eq{SXlargeN}
and therefore approaches the massless minimally coupled
gravitational mode at large rank. In the kinetically aligned
regime, the massive partner approaches the conformally coupled
mode and remains infrared regular.

\item {\it Chaotic horizon dynamics provides a candidate microscopic
mechanism for stochastic gravity.}

The soft mode couples to the microscopic horizon theory through
the force conjugate to the breathing coordinate. The Gaussian
horizon theory exhibits substantial force fluctuations but retains
coherent memory and therefore does not by itself generate
irreversible stochastic motion. Interacting many-body spectral
mixing is expected to produce a relaxing force correlator with
finite low-frequency spectral weight. Under this assumption,
coarse graining over the fast horizon degrees of freedom leads
to Langevin dynamics, with noise and dissipation determined by
the microscopic force correlators.
Thus the stochastic source is supplied by the microscopic
horizon dynamics rather than introduced as an external
phenomenological ingredient. It is important to test the stochastic
description by establishing the required decay of force correlations
in the full interacting theory. 

\item {\it Stochastic diffusion generates bosonic coarse-grained
entropy.}

The same force fluctuations that generate Langevin noise
suppress off-diagonal histories of the soft mode in the
influence-functional description. In the semiclassical regime,
the Fokker--Planck probability distribution describes the
coarse-grained distribution of the reduced state of the soft mode.
For pure Brownian diffusion, its coarse-grained entropy
grows with time.
Under appropriate mixing and central-limit conditions, the
accumulated stochastic displacement approaches a Gaussian
distribution. Consequently,
\be
\Delta S_X^{\mathrm{cg}}(T)=\frac12{\cal L}_T, \qquad
{\cal L}_T:=\ln\frac{\sigma_X^2(T)}{\sigma_X^2(T_0)}.
\ee
The coefficient relating entropy growth to logarithmic
variance broadening is therefore fixed by Gaussian statistics
and is independent of the microscopic diffusion coefficient.

\item {\it The quantum entropy correction leads to a gravitational
screening proposal.}

The degenerate fermionic parton sector provides the background
microstate entropy $S_0=2N^2=S_{\mathrm{GH}}^{(0)}$.
The stochastic bosonic dynamics generates an additional
coarse-grained entropy. Neglecting changes in the fermionic
entropy and intersector correlations at the order of this
correction, we propose the total thermodynamic entropy
$ S_{\mathrm{th}}(T)=
S_0+\Delta S_{\mathrm{cg}}^{\mathrm{all}}(T)$.
We further conjecture that this thermodynamic entropy is
represented by the Gibbons--Hawking entropy of the
backreacted de Sitter geometry:
%c11 causal horizon
\be
S_{\rm th}(T) = S_{\rm GH} (T). 
\ee
Within the same Gaussian stochastic approximation and
the proposed
entropy--gravity identification, the screening of the
Hubble parameter is described by the beta function
\be
\frac{dg}{d{\cal L}_T} = -\frac12g^2.
\ee
As stochastic diffusion increases the soft-mode variance,
the gravitational coupling decreases. At fixed Newton's constant, this
would correspond to a reduction of the Hubble parameter and effective
cosmological constant, representing an infrared screening of de Sitter
curvature under the proposed entropy--gravity identification.

\item {\it Radiative lifting controls the duration of the
entropy-producing regime.}

If the classically massless soft mode acquires a positive
radiative curvature, its late-time Brownian dynamics crosses
over to an Ornstein--Uhlenbeck process. The variance and
coarse-grained entropy then approach finite stationary values.
For the radiative scaling
and the assumed momentum-relaxation scaling, 
a parametrically extended Brownian regime requires
\be
\nu >2.
\ee
The case of equality is marginal. A negative radiative
curvature instead produces an instability, so that the
quadratic stochastic theory admits no late-time stationary
distribution. The sign and magnitude of radiative lifting
therefore determine the duration and possible endpoint of
the stochastic entropy-producing regime.

\end{enumerate}

Taken together, these results suggest the microscopic
mechanism
\be
\text{horizon chaos}
\longrightarrow \text{force  noise}
\longrightarrow \text{soft-mode diffusion}
\longrightarrow \text{entropy growth}
\longrightarrow \text{dS screening}.
\ee
The construction offers a microscopic origin for the stochastic de
Sitter dynamics previously proposed at the gravitational level and a
robust mechanism for entropy generation. Its interpretation as
gravitational screening rests on the conjectured identification of the
quantum-corrected thermodynamic entropy with the Gibbons--Hawking
entropy of the backreacted de Sitter geometry.

These results suggest that the quantum-horizon matrix model captures
nontrivial aspects of infrared quantum gravity. Establishing the full
dynamical mechanism requires computing the interacting microscopic
force correlators and deriving the resulting gravitational
backreaction of the de Sitter geometry.

\section{Discussion and Outlook}
We now turn to several broader implications of the framework
developed above and to some open questions that naturally follow from it.

\subsection{Black-hole and de Sitter spacetimes from
  common microscopic dynamics}

Our construction suggests that the Schwarzschild exterior and de
Sitter interior may emerge as distinct low-energy
geometries of the same microscopic horizon system. The fuzzy-sphere
background and its collective fluctuations provide boundary data
which, together with the gravitational junction conditions and
appropriate boundary conditions on each side, may determine the bulk
geometries. In this picture, the bulk metric is an emergent low-energy
field rather than a fundamental matrix variable.

The same microscopic horizon fluctuations can produce different
gravitational responses. On the de Sitter side, the massless
collective mode undergoes stochastic diffusion, generating
coarse-grained entropy and the proposed entropic screening of
curvature. On the Schwarzschild side, where a corresponding flat
direction is not guaranteed, the fluctuations may instead produce a
dissipative radial response.
Thus, the two geometries share a common microscopic origin but exhibit
different infrared dynamics determined by their respective
gravitational constraints and boundary conditions. Deriving the bulk
gravitational equations and their stochastic responses directly from
the matrix model remains an important open problem.

\subsection{Stochastic dynamics on the black-hole side}

The same microscopic force fluctuations may also induce stochastic
fluctuations of the Schwarzschild horizon. Determining their dynamics
requires identifying the Schwarzschild gravitational response coordinate
and its coupling to the microscopic breathing mode.

Define the fractional Schwarzschild horizon fluctuation by
\be
\varpi_{\rm BH}:=\frac{\delta R_{\rm BH}}{R_{\rm BH}}.
\ee
If its leading gravitational response agrees with the microscopic
breathing fluctuation,
\be\label{varpi-BH}
\varpi_{\rm BH}=\varpi+\cdots,
\ee
then the corresponding variations satisfy
\be
\frac{\delta A}{A}=2\varpi_{\rm BH},\qquad
\frac{\delta M}{M}=\varpi_{\rm BH},\qquad
\frac{\delta T_H}{T_H}=-\varpi_{\rm BH}.
\ee
This identification is an additional assumption, independent of
the de Sitter field dictionary.
The resulting fluctuations also affect the Hawking luminosity. At
leading Schwarzschild order,
\be
L_H\propto A T_H^4\propto R_{\rm BH}^{-2},
\ee
giving
\be
\frac{\delta L_H}{L_H}=-2\varpi_{\rm BH},
\ee
up to corrections from the frequency dependence of greybody
factors. Thus, horizon-radius fluctuations are anticorrelated with
fluctuations of the Hawking temperature and luminosity. Computing
their unequal-time correlations would provide a direct probe of
microscopic horizon dynamics through Hawking radiation.

This picture complements semiclassical stochastic gravity, where
metric fluctuations are driven by matter stress-tensor fluctuations
through the Einstein--Langevin equation \cite{Hu:2008rga}. Here the
horizon matrix model provides an intrinsic source of quantum-geometric
noise. It would be interesting to derive the Schwarzschild response
and compare the resulting noise and dissipation kernels with those
generated by quantum matter fluctuations in semiclassical stochastic
gravity.

\subsection{Quantum state of the common horizon}

The stochastic coefficients depend not only on the Hamiltonian but
also on the quantum state of the horizon. This issue is especially
important for a common black-hole--de Sitter horizon because the two
bulk regions come with different natural notions of time, positive
frequency, and thermal state. The junction conditions determine the
background geometry and the relative normalization of the
corresponding time coordinates, but they do not by themselves
determine the quantum state.

In the main text, we have employed an equilibrium KMS state
only as a benchmark.
A more general stationary state need not obey the equilibrium KMS
relation. An Unruh-like state is a natural possibility when one side
of the horizon is associated with outgoing Hawking radiation, while
more general nonequilibrium stationary states may also be compatible
with the soldered geometry. In such cases the symmetrized noise kernel
and the retarded response are independent dynamical quantities.
For a general nonequilibrium state, those quantities, together with
the momentum-relaxation time and onset of the Brownian regime, must be
computed from the state-dependent correlators.

More fundamentally, the soldered Schwarzschild--de Sitter geometry may
be viewed as two gravitational open systems coupled through a common
dynamical quantum boundary. The horizon is not merely a geometric
interface on which boundary conditions are imposed; it carries its own
microscopic quantum degrees of freedom, quantum state, fluctuations, and
response. From this perspective, the appropriate quantum state of the
de Sitter region need not be specified independently of the black-hole
exterior or of the horizon degrees of freedom that couple the two
regions. In particular, the notion of a de Sitter ``vacuum'' may
differ from that obtained by quantizing an isolated de Sitter
spacetime and imposing the Bunch--Davies state. The problem of vacuum
selection is therefore naturally promoted to a dynamical question for
the coupled black-hole--horizon--de Sitter system as a whole.

\appendix
\section{Gaussian Force Correlator and Zero-Frequency Spectral Support}

In this appendix, we collect the Gaussian calculation underlying the
force estimates used in Sec.~4. The purpose is to establish three
properties of the microscopic force conjugate to the breathing mode:
its equal-time variance, its characteristic frequency bandwidth, and
the absence of regular relaxing spectral weight at zero frequency. The
last point is crucial for distinguishing coherent Gaussian
fluctuations from stochastic noise. All quantities in this appendix
are evaluated in the microscopic $q_{10}$ normalization; the projection
onto the massless collective direction is performed in Sec.~4.

We consider the stable Gaussian
fluctuation modes of the fuzzy sphere. Denoting a mode collectively by
$n=(L,K,Q)$, their quadratic frequencies are
\be \omega_n=\Omega_T\sqrt{k_n},
\ee
where $k_n>0$ denotes the corresponding dimensionless classical
curvature and the prime below indicates that nonphysical modes are
omitted. Differentiating the quadratic Hamiltonian with respect to the
microscopic breathing coordinate gives
\be F_{10}^{(2)}=-\frac{\Omega_T^2}{\sqrt{D_N}}\sum_n'(k_n+2)Y_n^2 .
\ee
Subtracting the expectation value and applying Gaussian Wick
contractions gives the connected vacuum correlator
\be
C_{F_{10}}^{\rm free}(T)=\frac{\Omega_T^2}{2D_N}\sum_n'
\frac{(k_n+2)^2}{k_n}\cos(2\omega_nT).
\ee
The corresponding symmetrized spectral density is
\be
S_{F_{10}}^{\rm free}(\omega)=\frac{\pi\Omega_T^2}{2D_N}\sum_n'
\frac{(k_n+2)^2}{k_n}\left[\delta(\omega-2\omega_n)+\delta(\omega+2\omega_n)
  \right].
\ee
Thus the Gaussian force spectrum consists of discrete pair lines. The
force fluctuations can be large even though the spectrum contains no
smooth low-frequency continuum.

To determine the large-$N$ scaling of
the equal-time variance, it is useful to organize the sum into
angular-momentum multiplets. For the two physical branches at large
angular momentum,
\be
k_{LK}=L^2\left[1+O(L^{-1})\right],\qquad d_{LK}=2L\left[1+O(L^{-1})\right].
\ee
We define the force weight of a multiplet by
\be
W_{LK}=d_{LK}\frac{(k_{LK}+2)^2}{k_{LK}}.
\ee
Summing over the two physical branches at fixed $L$ gives
\be W_L^{\rm tot}=4L^3\left[1+O(L^{-1})\right].
\ee
The force variance is therefore strongly ultraviolet dominated,
\be \sum_LW_L^{\rm tot}=O(N^4).
\ee
Using
\be D_N=\frac{N(N^2-1)}{4}=O(N^3),
\ee
together with the de Sitter-time microscopic scale
\be \Omega_T=\frac{3H}{\pi N}\left[1+O(N^{-1})\right],
\ee
we obtain
\be \label{CC1}
C_{F_{10}}^{\rm free}(T=0)\sim\frac{H^2}{N}.
\ee

The ultraviolet dominance is particularly strong. At leading order,
the fraction of the force weight carried by modes with angular momentum
larger than half the cutoff is
\be \frac{\displaystyle\int_{N/2}^{N}dL\,L^3}{\displaystyle\int_0^{N}dL\,L^3}
=\frac{15}{16}.
\ee
The force therefore probes predominantly microscopic modes near the
fuzzy-sphere ultraviolet cutoff.
The characteristic force bandwidth can be quantified by the force-weighted
second spectral moment,
\be (\Delta\omega_{F_{10}})^2
:=\frac{\displaystyle\sum_{L,K}'W_{LK}(2\omega_{LK})^2}
    {\displaystyle\sum_{L,K}'W_{LK}}.
    \ee
At large $N$ the asymptotic sum gives
\be (\Delta\omega_{F_{10}})^2
=\frac{8}{3}\Omega_T^2N^2\left[1+O(N^{-1})\right], \ee
and hence
\be \label{CC2}
\Delta\omega_{F_{10}}=O(H).
\ee
The same scale appears directly in the time-domain correlator. Introducing
\be x=\frac{L}{N},\qquad z=2\Omega_TNT, \ee
the leading large-$N$ result becomes
\be \frac{C_{F_{10}}^{\rm free}(T)}{C_{F_{10}}^{\rm free}(0)}\simeq4\int_0^1dx\,x^3
\cos(zx). \ee
The first zero occurs at
\be z\simeq1.956,
\ee
so the characteristic Gaussian dephasing time is
\be T_{\rm deph}=O\!\left[(N\Omega_T)^{-1}\right]=O(H^{-1}).
\ee
This decrease of the correlator is produced by destructive
interference among many discrete spectral lines. It should not be
interpreted as irreversible relaxation. At finite $N$ the Gaussian
evolution remains coherent and can exhibit recurrences.  The
distinction becomes especially important at zero frequency.

Since
every retained stable Gaussian oscillator has nonzero frequency, the
pair spectrum contains no regular contribution at the origin,
\be S_{F_{10},{\rm pair}}^{\rm free}(\omega=0)=0.
\ee
A stationary Gaussian ensemble may in addition contain fluctuations of
the oscillator occupation numbers. Since the individual occupation
numbers are conserved in the free Gaussian theory, these fluctuations
produce a time-independent contribution to the connected force
correlator,
\be
C_{F_{10},{\rm occ}}^{\rm free}(T)=A_{F_{10}},
\ee
and therefore a singular zero-frequency term,
\be
S_{F_{10},{\rm occ}}^{\rm free}(\omega)=2\pi A_{F_{10}}\delta(\omega).
\ee
This contribution represents a nondecaying conserved component of the
force in the free theory. It has infinite memory and is not regular
stochastic noise. Once interactions are included, the individual
Gaussian occupation numbers are no longer generally conserved, and
their nonconserved combinations can relax and contribute to the regular
low-frequency force spectrum. Any exactly conserved component must
still be projected out before defining the relaxing force spectrum
relevant to the Langevin description. Accordingly, throughout the
main text the notation $S_{F_{10}}(0)$ refers to the finite regular
limit of the relaxing projected force spectrum rather than to a
possible delta function at the origin. With this definition the
Gaussian theory gives
\be \label{CC3}
S_{F_{10},{\rm reg}}^{\rm free}(0)=0.
\ee

We therefore arrive at the two complementary Gaussian results
\eq{CC1},\eq{CC2},
while the regular stochastic spectral weight \eq{CC3}
remains absent.
The free horizon theory thus contains substantial microscopic force
fluctuations and an order-$H$ frequency bandwidth, but these
fluctuations remain coherent. A nonzero smooth zero-frequency spectrum
requires interactions to mix the microscopic excitations and relax the
nonconserved part of the force. This is the additional dynamical input
assumed in the stochastic description developed in Sec.~4.

\section*{Acknowledgments}

This paper is dedicated to the memory of Yoshihisa Kitazawa. I had the
pleasure of discussing his work on infrared quantum effects in de
Sitter space with him on several occasions. His deep insights into de
Sitter infrared dynamics have remained an important inspiration for
the present work. I remember him with warmth, admiration, and
heartfelt gratitude.
%c10
I also thank Dimitris Giataganas, Bei-Lok Hu, Hikaru Kawai,
Robert de Mello Koch, Gary Shiu
and the participants of the East Asia Joint Workshop on Fields and
Strings 2026 and Frontiers in Theoretical Physics 2026, held at KIAS,
for stimulating discussions and valuable comments on this work.
Support from the National Center for Theoretical
Sciences (NCTS), the National Science and Technology Council of Taiwan
under Grant No. 113-2112-M-007-039-MY3, and the National Tsing Hua
University 2025 Talent Development Fund through a TSAI WANG, YUAN-YANG
Distinguished Talent Chair Professorship is gratefully acknowledged.

\bibliographystyle{utphys}
\bibliography{references}   

@article{Christie:2026dwx,
    author = "Christie, Robson and Joo, Jaewoo and Kaplanek, Greg and Vennin, Vincent and Wands, David",
    title = "{Quantum Stochastic Inflation}",
    eprint = "2606.12636",
    archivePrefix = "arXiv",
    primaryClass = "hep-th",
    month = "6",
    year = "2026"
}

@article{Li:2026lwl,
    author = "Li, Yue-Zhou",
    title = "{Stochastic inflation as an open quantum system II: open effective field theory and stochastic matching}",
    eprint = "2605.21929",
    archivePrefix = "arXiv",
    primaryClass = "hep-th",
    month = "5",
    year = "2026"
}

@article{Hu:1991di,
    author = "Hu, B. L. and Paz, Juan Pablo and Zhang, Yu-hong",
    title = "{Quantum Brownian motion in a general environment:  Exact master equation with nonlocal dissipation and colored noise}",
    reportNumber = "FPRINT-91-19",
    doi = "10.1103/PhysRevD.45.2843",
    journal = "Phys. Rev. D",
    volume = "45",
    pages = "2843--2861",
    year = "1992"
}

@article{Kubo:1966fyg,
    author = "Kubo, R.",
    title = "{The fluctuation-dissipation theorem}",
    doi = "10.1088/0034-4885/29/1/306",
    journal = "Rept. Prog. Phys.",
    volume = "29",
    number = "1",
    pages = "255",
    year = "1966"
}

@article{Grabert:1988yt,
    author = "Grabert, H. and Schramm, P. and Ingold, G. L.",
    title = "{Quantum Brownian motion: The Functional integral approach}",
    doi = "10.1016/0370-1573(88)90023-3",
    journal = "Phys. Rept.",
    volume = "168",
    pages = "115--207",
    year = "1988"
}

@article{Polyakov:2007mm,
    author = "Polyakov, A. M.",
    title = "{De Sitter space and eternity}",
    eprint = "0709.2899",
    archivePrefix = "arXiv",
    primaryClass = "hep-th",
    reportNumber = "PUPT-2244",
    doi = "10.1016/j.nuclphysb.2008.01.002",
    journal = "Nucl. Phys. B",
    volume = "797",
    pages = "199--217",
    year = "2008"
}

@article{Polyakov:2009nq,
    author = "Polyakov, A. M.",
    title = "{Decay of Vacuum Energy}",
    eprint = "0912.5503",
    archivePrefix = "arXiv",
    primaryClass = "hep-th",
    reportNumber = "PUPT-2320",
    doi = "10.1016/j.nuclphysb.2010.03.021",
    journal = "Nucl. Phys. B",
    volume = "834",
    pages = "316--329",
    year = "2010"
}

@article{Aalsma:2019rpt,
    author = "Aalsma, Lars and Parikh, Maulik and Van Der Schaar, Jan Pieter",
    title = "{Back(reaction) to the Future in the Unruh-de Sitter State}",
    eprint = "1905.02714",
    archivePrefix = "arXiv",
    primaryClass = "hep-th",
    doi = "10.1007/JHEP11(2019)136",
    journal = "JHEP",
    volume = "11",
    pages = "136",
    year = "2019"
}

@article{Aalsma:2020aib,
    author = "Aalsma, Lars and Shiu, Gary",
    title = "{Chaos and complementarity in de Sitter space}",
    eprint = "2002.01326",
    archivePrefix = "arXiv",
    primaryClass = "hep-th",
    reportNumber = "MAD-TH-20-01",
    doi = "10.1007/JHEP05(2020)152",
    journal = "JHEP",
    volume = "05",
    pages = "152",
    year = "2020"
}

@article{Kitamoto:2019rij,
    author = "Kitamoto, Hiroyuki and Kitazawa, Yoshihisa and Matsubara, Takahiko",
    title = "{de Sitter duality and logarithmic decay of dark energy}",
    eprint = "1908.02534",
    archivePrefix = "arXiv",
    primaryClass = "hep-th",
    reportNumber = "KEK-Cosmo-245, KEK-TH-2142, NCTS-TH/1906, NCTS-TH-1906",
    doi = "10.1103/PhysRevD.101.023504",
    journal = "Phys. Rev. D",
    volume = "101",
    number = "2",
    pages = "023504",
    year = "2020"
}

@article{Kitazawa:2020oaf,
    author = "Kitazawa, Yoshihisa",
    title = "{Holographic {\ensuremath{\beta}} Function in de Sitter Space}",
    eprint = "2011.14640",
    archivePrefix = "arXiv",
    primaryClass = "hep-th",
    reportNumber = "KEK-TH- 2223",
    doi = "10.7566/JPSJ.91.074003",
    journal = "J. Phys. Soc. Jap.",
    volume = "91",
    number = "7",
    pages = "074003",
    year = "2022"
}

@article{Kitazawa:2024deb,
    author = "Kitazawa, Yoshihisa",
    title = "{Fokker-Planck Equation and de Sitter Duality}",
    eprint = "2406.00886",
    archivePrefix = "arXiv",
    primaryClass = "hep-th",
    reportNumber = "KEK-TH-2613",
    month = "6",
    year = "2024"
}

@article{Garriga:2007zk,
    author = "Garriga, Jaume and Tanaka, Takahiro",
    title = "{Can infrared gravitons screen Lambda?}",
    eprint = "0706.0295",
    archivePrefix = "arXiv",
    primaryClass = "hep-th",
    reportNumber = "KUNS-2076",
    doi = "10.1103/PhysRevD.77.024021",
    journal = "Phys. Rev. D",
    volume = "77",
    pages = "024021",
    year = "2008"
}

@article{Tsamis:2007is,
    author = "Tsamis, N. C. and Woodard, R. P.",
    title = "{Comment on `Can infrared gravitons screen Lambda?'}",
    eprint = "0708.2004",
    archivePrefix = "arXiv",
    primaryClass = "hep-th",
    reportNumber = "CRETE-07-11, UFIFT-QG-07-02",
    doi = "10.1103/PhysRevD.78.028501",
    journal = "Phys. Rev. D",
    volume = "78",
    pages = "028501",
    year = "2008"
}

@article{Higuchi:2011vw,
    author = "Higuchi, Atsushi and Marolf, Donald and Morrison, Ian A.",
    title = "{de Sitter invariance of the dS graviton vacuum}",
    eprint = "1107.2712",
    archivePrefix = "arXiv",
    primaryClass = "hep-th",
    doi = "10.1088/0264-9381/28/24/245012",
    journal = "Class. Quant. Grav.",
    volume = "28",
    pages = "245012",
    year = "2011"
}

@article{Kitamoto:2012vj,
    author = "Kitamoto, Hiroyuki and Kitazawa, Yoshihisa",
    title = "{Soft Graviton effects on Gauge theories in de Sitter Space}",
    eprint = "1204.2876",
    archivePrefix = "arXiv",
    primaryClass = "hep-th",
    reportNumber = "KEK-TH-1539",
    doi = "10.1103/PhysRevD.87.124004",
    journal = "Phys. Rev. D",
    volume = "87",
    number = "12",
    pages = "124004",
    year = "2013"
}

@article{Kitamoto:2014gva,
    author = "Kitamoto, Hiroyuki and Kitazawa, Yoshihisa",
    title = "{Time Dependent Couplings as Observables in de Sitter Space}",
    eprint = "1402.2443",
    archivePrefix = "arXiv",
    primaryClass = "hep-th",
    reportNumber = "KEK-TH-1704, SNUTP14-003",
    doi = "10.1142/S0217751X14300166",
    journal = "Int. J. Mod. Phys. A",
    volume = "29",
    number = "8",
    pages = "1430016",
    year = "2014"
}

@article{Kitamoto:2016yfy,
    author = "Kitamoto, Hiroyuki and Kitazawa, Yoshihisa",
    title = "{Effective Lagrangian in de Sitter Spacetime}",
    eprint = "1609.03293",
    archivePrefix = "arXiv",
    primaryClass = "hep-th",
    reportNumber = "KEK-TH-1933, KUNS-2641",
    doi = "10.1103/PhysRevD.95.025017",
    journal = "Phys. Rev. D",
    volume = "95",
    number = "2",
    pages = "025017",
    year = "2017"
}

@article{Hu:2008rga,
    author = "Hu, B. L. and Verdaguer, E.",
    title = "{Stochastic Gravity: Theory and Applications}",
    eprint = "0802.0658",
    archivePrefix = "arXiv",
    primaryClass = "gr-qc",
    doi = "10.12942/lrr-2008-3",
    journal = "Living Rev. Rel.",
    volume = "11",
    pages = "3",
    year = "2008"
}

@article{Starobinsky:1982ee,
    author = "Starobinsky, Alexei A.",
    title = "{Dynamics of Phase Transition in the New Inflationary Universe Scenario and Generation of Perturbations}",
    doi = "10.1016/0370-2693(82)90541-X",
    journal = "Phys. Lett. B",
    volume = "117",
    pages = "175--178",
    year = "1982"
}

@article{Starobinsky:1986fx,
    author = "Starobinsky, Alexei A.",
    title = "{STOCHASTIC DE SITTER (INFLATIONARY) STAGE IN THE EARLY UNIVERSE}",
    doi = "10.1007/3-540-16452-9_6",
    journal = "Lect. Notes Phys.",
    volume = "246",
    pages = "107--126",
    year = "1986"
}

@article{Burgess:2014eoa,
    author = "Burgess, C. P. and Holman, R. and Tasinato, G. and Williams, M.",
    title = "{EFT Beyond the Horizon: Stochastic Inflation and How Primordial Quantum Fluctuations Go Classical}",
    eprint = "1408.5002",
    archivePrefix = "arXiv",
    primaryClass = "hep-th",
    reportNumber = "CERN-PH-TH-2014-142",
    doi = "10.1007/JHEP03(2015)090",
    journal = "JHEP",
    volume = "03",
    pages = "090",
    year = "2015"
}

@article{Collins:2017haz,
    author = "Collins, Hael and Holman, Richard and Vardanyan, Tereza",
    title = "{The quantum Fokker-Planck equation of stochastic inflation}",
    eprint = "1706.07805",
    archivePrefix = "arXiv",
    primaryClass = "hep-th",
    doi = "10.1007/JHEP11(2017)065",
    journal = "JHEP",
    volume = "11",
    pages = "065",
    year = "2017"
}

@article{Li:2025azq,
    author = "Li, Yue-Zhou",
    title = "{Stochastic Inflation as an Open Quantum System}",
    eprint = "2507.02070",
    archivePrefix = "arXiv",
    primaryClass = "hep-th",
    doi = "10.1103/hlfz-8lhg",
    journal = "Phys. Rev. Lett.",
    volume = "136",
    number = "7",
    pages = "071501",
    year = "2026"
}

@book{IbragimovLinnik1971,
  author    = {I. A. Ibragimov and Yu. V. Linnik},
  title     = {Independent and Stationary Sequences of Random Variables},
  publisher = {Wolters-Noordhoff},
  address   = {Groningen, The Netherlands},
  year      = {1971},
  isbn      = {9001418856},
  note      = {Chapter 18, especially Section 18.7,
               ``The Central Limit Theorem in Continuous Time'',
               pp. 362--364}
}

@article{Thorne:1980rmp,
  author        = "Thorne, Kip S.",
  title         = "{Multipole Expansions of Gravitational Radiation}",
  journal       = "Reviews of Modern Physics",
  volume        = "52",
  number        = "2",
  pages         = "299--339",
  year          = "1980",
  doi           = "10.1103/RevModPhys.52.299",
  publisher     = "American Physical Society"
}

@article{Chu:2026rac,
    author = "Chu, Chong-Sun",
    title = "{Spectral Chaos from a Fluctuating Quantum Horizon}",
    eprint = "2609.15286",
    archivePrefix = "arXiv",
    primaryClass = "hep-th",
    month = "9",
    year = "2026"
}

@article{Chu:2026dhx,
    author = "Chu, Chong-Sun",
    title = "{Quantum Stability of the Fuzzy Sphere Black Hole Horizon}",
    eprint = "2608.16374",
    archivePrefix = "arXiv",
    primaryClass = "hep-th",
    month = "8",
    year = "2026"
}

@article{Chu:2026qom,
    author = "Chu, Chong-Sun",
    title = "{From Horizon Microstates to the Black Hole Membrane}",
    eprint = "2607.28746",
    archivePrefix = "arXiv",
    primaryClass = "hep-th",
    month = "7",
    year = "2026"
}

@article{Chu:2026vzj,
    author = "Chu, Chong-Sun",
    title = "{Quantum Horizon and Quantum Membrane Paradigm from Black Hole Quantum Mechanics}",
    eprint = "2607.10561",
    archivePrefix = "arXiv",
    primaryClass = "hep-th",
    month = "7",
    year = "2026"
}

@article{Chu:2024qil,
    author = "Chu, Chong-Sun",
    title = "{Matrix model proposal for quantum gravity and the quantum mechanics of black holes}",
    eprint = "2406.01466",
    archivePrefix = "arXiv",
    primaryClass = "hep-th",
    doi = "10.1103/l7cq-tx4g",
    journal = "Phys. Rev. D",
    volume = "112",
    number = "6",
    pages = "066001",
    year = "2025"
}

@article{Chu:2024edh,
    author = "Chu, Chong-Sun",
    title = "{Quantum Kerr black hole from matrix theory of quantum gravity}",
    eprint = "2406.12704",
    archivePrefix = "arXiv",
    primaryClass = "hep-th",
    doi = "10.1103/5rwl-tr4y",
    journal = "Phys. Rev. D",
    volume = "112",
    number = "4",
    pages = "046014",
    year = "2025"
}

@article{Chu:2026wyh,
    author = "Chu, Chong-Sun",
    title = "{Hawking Radiation from Tunneling in Black Hole Quantum Mechanics}",
    eprint = "2603.12199",
    archivePrefix = "arXiv",
    primaryClass = "hep-th",
    month = "3",
    year = "2026"
}

@article{Chu:2026kqt,
    author = "Chu, Chong-Sun",
    title = "{Quantum Horizon Tadpole and Emergence of a de Sitter Interior}",
    eprint = "2608.21178",
    archivePrefix = "arXiv",
    primaryClass = "hep-th",
    month = "8",
    year = "2026"
}
\end{document}